\pdfoutput=1
\documentclass[a4paper,12pt]{article}
\usepackage[utf8]{inputenc}
\usepackage{amssymb,amsmath,amsfonts,bbm,dsfont,graphicx,hyphenat,bm,float,a4wide,xcolor}
\usepackage[english]{babel}
\usepackage[colorlinks,linkcolor=blue,citecolor=blue,urlcolor=blue]{hyperref}
\usepackage[square,numbers,sort&compress]{natbib}
\usepackage[caption=false]{subfig}
\numberwithin{equation}{section}

\newcommand{\beq}{\begin{eqnarray}}
\newcommand{\eeq}{\end{eqnarray}}
\newcommand{\non}{\nonumber\\}

\DeclareMathOperator{\Og}{O}
\DeclareMathOperator{\SO}{SO}

\newcommand{\p}{\partial}
\renewcommand{\i}{\mathrm{i}}
\renewcommand{\d}{\mathop{}\!\mathrm{d}}

\DeclareMathOperator{\diag}{\rm diag}

\DeclareMathOperator{\eom}{eom}

\newcommand{\tR}{\widetilde{R}}

\newcommand{\bea}{\begin{eqnarray}}
\newcommand{\eea}{\end{eqnarray}}
\newcommand{\be}{\begin{equation}}
\newcommand{\ee}{\end{equation}}

\def\E{{\cal E}}

\begin{document}
\title{
\vskip 20pt
\bf{Higher-dimensional magnetic Skyrmions}
}
\vskip 40pt  

\author{
Sven Bjarke Gudnason,$^{1,}$\thanks{gudnason@henu.edu.cn} \ \  Stefano Bolognesi,$^{2,}$\thanks{stefano.bolognesi@unipi.it} \ \ Roberto Menta$^{3,}$\thanks{roberto.menta@sns.it}
\\[13pt]
   {\footnotesize	$^{1}$Institute of Contemporary Mathematics, School of Mathematics and Statistics,
  }\\[-5pt]
  {\footnotesize
    Henan University, Kaifeng, Henan 475004, P.~R.~China
  }\\[2pt] 
  {\footnotesize
    $^{2}$Department of Physics ``E. Fermi", University of Pisa, and INFN, Sezione di Pisa}\\[-5pt]
  {\footnotesize
    Largo Pontecorvo, 3, Ed. C, 56127 Pisa, Italy}\\[2pt]
  {\footnotesize
    $^{3}$Scuola Normale Superiore, Piazza dei Cavalieri, 7, and Laboratorio NEST, 
    }\\[-5pt]
  {\footnotesize
    Piazza S. Silvestro, 12, 56127  Pisa, Italy}
  \\[1pt]
}

\date{\small\today}

\vskip 8pt
\maketitle

\begin{abstract}
We propose a generalization of the theory of magnetic Skyrmions in
chiral magnets in two dimensions to a higher-dimensional theory with
magnetic Skyrmions in three dimensions and an $S^3$ target space,
requiring a 4-dimensional magnetization vector.
A physical realization of our theory could be made using
a synthetic dimension, recently promoted and realized in condensed
matter physics.
In the simplest incarnation of the theory, we find a Skyrmion and a
sphaleron -- the latter being an unstable soliton.
Including also the Skyrme term in the theory enriches the spectrum to a
small metastable Skyrmion, an unstable sphaleron and a large stable
Skyrmion.
\end{abstract}

\newpage
\tableofcontents

\section{Introduction}\label{intro}

The magnetic Skyrmion is gaining recognition as a prominent soliton in 2-dimensional magnetic physics, particularly in the context of chiral magnets and thin films~\cite{1989JETP...68..101B,1995JETPL..62..247B,Nagaosa2013,doi:10.1126/science.1166767,Yu2010}, as well as spintronics and quantum technologies~\cite{fert2017magnetic, everschor2018perspective, luo2021skyrmion, psaroudaki2023skyrmion}.
The name Skyrmion comes from the 3-dimensional soliton in the Skyrme
model, proposed by T.~Skyrme in the 1960s as a model of the
nucleus~\cite{skyrme1961non,skyrme1962unified}, further validated as a
low-energy effective field theory
of QCD at large $N_c$ (number of colors)
by E.~Witten~\cite{Witten:1983tw,Witten:1983tx} and later used as a validation of the Sakai-Sugimoto model as evidence that the model really describes QCD at low energies~\cite{Sakai:2004cn}.
The Skyrmion in 2 or 3 dimensions is a texture soliton, which differs
from a defect-type soliton by mapping the entire 2- or 3-dimensional
space to the target space. In particular, a magnetic Skyrmion in two
dimensions is mapping the entire plane to an $S^2$ target space, with
the latter being the normalized magnetization vector $\bm{n}=(n^1,n^2,n^3)$ such that $\bm{n}\cdot\bm{n}=1$~\cite{roessler2006spontaneous,
  Rossler:2010st, Ezawa:2010uy, Banerjee:2014hna, melcher2014chiral,
  Rybakov:2018bxt, schroers2019gauged, barton2020magnetic,
  kuchkin2020magnetic, ross2021skyrmion, hill2021chiral,
  schroers2021solvable, Amari:2022boe, Hanada:2023lnm}.
It differs drastically from the 3-dimensional
Skyrmion by the fact that: 1) it is stabilized by a
\emph{lower}-dimensional operator (i.e.~less derivatives than
number of spatial dimensions) and 2) it is chiral.
The latter implies that either (isolated) magnetic Skyrmions
stably exist or the corresponding anti-Skyrmions, but not both of
them.\footnote{For isolated axially symmetric Skyrmions, this
is the case and the stable (anti-)Skyrmions are those corresponding to
a negative DMI energy (with its coefficient).
However, both Skyrmions and anti-Skyrmion may exist for certain
potentials and on domain lines, see e.g.~\cite{PhysRevB.101.064408,kuchkin2020magnetic}.
}
This is different from 3-dimensional Skyrmions, where a
higher-dimensional operator (i.e.~more derivatives than number of
spatial dimensions) is used to stabilize the soliton -- called a
Skyrmion -- against collapse.
Indeed the higher-derivative (or smaller-derivative) terms are a way to avoid
Derrick's theorem \cite{Derrick:1964ww}. 

In this paper, we consider the generalization of the 2-dimensional
magnetic Skyrmion with target space $S^2$ to a 3-dimensional magnetic
Skyrmion\footnote{A preliminary investigation
started in Ref.~\cite{menta2023magnetic}.} with target space $S^3$.\footnote{
  Solitons in $^3$He-B phase also have a 3-sphere for a target space
  as well as a spin-orbit interaction, but to the best of our
  knowledge does not give rise to a DMI term
  \cite{Volovik:1977,10.1093/acprof:oso/9780199564842.001.0001}.}
Since the unit magnetization vector in three dimensions gives rise to
$S^2$, we need a unit magnetization vector in \emph{four}
dimensions.\footnote{We use the notation ``magnetization vector'' for
the 4-dimensional generalization of the 3-dimension magnetization vector,
although it is physically different. For example, it is not possible
to write $\bm{B}\cdot\bm{n}$, since the 3-dimensional magnetic field
has 3 components and a would-be 4-dimensional magnetic field cannot be
dualized to a vector (the 4-dimensional Hodge dual of a 2-form is also
a 2-form).}
We will henceforth denote the construct as a higher-dimensional
magnetic Skyrmion. 

The physical realization of a 3-dimensional magnetic Skyrmion
could be made by
the incorporation of a synthetic (extra) dimension, as has
been demonstrated in a number of physical systems -- e.g.~see
Refs.~\cite{Boada2012,Celi2014,Ozawa2015,Price2015,Yuan2018,Ozawa2019,Wang2020,Price2022,Hazzard2023,Arguello-Luengo2024}.
Until now, the 3-dimensional counterpart of 2D magnetic Skyrmions
were magnetic Hopfions, 3-dimensional solitons living in a magnetization vector that also lives in three dimensions.
Recent numerical computations
\cite{sutcliffe2017skyrmion,sutcliffe2018hopfions,sutcliffe2018hopfionsReview,tai2018static,liu2018binding}
suggest that similar objects may exist at the nanoscale in frustrated
and chiral magnets.
Micrometer-sized Hopfions have recently been created and observed experimentally in chiral
ferromagnetic fluids \cite{ackerman2017static,kent2021creation}. The
3D magnetic Skyrmions which we propose, instead, have $S^3$ as a target
space, the same as used in the standard Skyrme model in three
dimensions.

In this paper, we concentrate on the spherically symmetric case which
enjoys $\SO(3)_s$ invariance, with the subscript $s$ denoting space. We take the generalized
Dzyaloshinskii-Moriya interaction (DMI)~\cite{dzyaloshinsky1958thermodynamic, moriya1960anisotropic, moriya1960new} to have only one derivative of
the magnetization vector and a single component of the same vector.
In four dimensions, this means that such a term must be contracted with
a tensor that possesses two $\SO(4)$-indices and a single spatial
index: we denote such a tensor by $\Theta$.
$\SO(4)$ invariance dictates that $\Theta$ is antisymmetric in
$\SO(4)$-indices and we use this fact, as well as global rotations to
rotate the tensor to its \emph{standard form}.
We further impose $\SO(3)_{\rm diag}$ invariance, where we are locking spatial
rotations with an $\SO(3)_t\subset\SO(4)$ rotation, in order to obtain a
spherically symmetric soliton (the subscript $t$ denotes the target
  space and 'diag' denotes the locked combined symmetry).
This procedure leads to \emph{two} different invariant structures in
the $\Theta$ tensor.
However, only one of these two structures remains nonvanishing when a
spherically symmetric Ansatz is imposed. 
We furthermore show that both the Bloch-type and the N\'eel-type DMI
generalizations lead to the same term when $\SO(3)_{\rm diag}$ invariance is
imposed. 

A special feature of this higher-dimensional magnetic Skyrme model, is
that it has multiple solutions. Indeed in its simplest incarnation, it
has a Skyrmion and a sphaleron \cite{manton2019inevitability}, with
the latter being unstable and much smaller than the Skyrmion.
Finally, we further extend the model by including the standard $S^3$
Skyrme term, which further enriches the phase diagram of the model, by
engendering the model with a small metastable Skyrmion, an unstable
sphaleron and a stable Skyrmion -- at least in a finite region of the
model's parameter space.

The paper is organized as follows.
In Sec.~\ref{sec:3D}, we consider the minimal generalization of 2D
magnetic Skyrmions to 3D magnetic Skyrmions with a 4-dimensional
magnetization vector.
In particular, in Sec.~\ref{sec:genDM} we consider the generalizations
of the DM term and its reductions due to symmetry considerations.
In Sec.~\ref{sec:sph_DM_Sk}, we then study spherically symmetric
Skyrmions numerically and consider the limit of the kinetic term being
absent, leading to the so-called restricted model (analogous to other
Skyrme-type models).
In Sec.~\ref{sec:sphalerons}, we then turn to unstable solitons --
the sphalerons of the theory -- which are the smaller ones and harder to find.
In Sec.~\ref{sec:magSkHopf}, we describe a limit with an additional
potential, in which the model eventually reduces to a chiral magnetic
Hopfion model.
In Sec.~\ref{sec:hybrid}, we then consider including the Skyrme term which
enriches the theory with an extra and tiny Skyrmion that is only
metastable.
In Sec.~\ref{sec:synthetic}, we discuss the possibility of using a
synthetic dimension to realize our theory in the real world.
We conclude in Sec.~\ref{sec:conclusion} with a discussion and
conclusion.

\section{Higher-dimensional magnetic Skyrmions}\label{sec:3D}

A possible extension of planar magnetic Skyrmions can be realized by
increasing the dimension of the base space and target space from two
to three dimensions~\cite{menta2023magnetic}.
This amounts to a magnetization unit vector that would live in a
4-dimensional Cartesian space. For a physical realization, advanced
concepts such as synthetic dimensions must be considered, we will deal with
this argument in Sec.~\ref{sec:synthetic}. 
In symbols, the magnetization field is now a map
$\bm{n} : \mathbb{R}^3 \to S^3$ where $\bm{n}=(n^1,n^2,n^3,n^4)$ and
$\bm{n}\cdot\bm{n}=1$.
Let us consider the energy functional of the following form:
\beq
E[\bm{n}] = \sum_{\ell=0,1,2} E_{\ell}[\bm{n}] \ , \qquad
E_{\ell}[\bm{n}]  = \int\d^3x \; \E_{\ell}(\bm{n})\ .
\label{Emagsky}
\eeq
The Heisenberg exchange term is
\beq
\E_2(\bm{n}) = \frac{1}{2} \partial_i\bm{n}\cdot\partial_i\bm{n}\ .
\eeq
As potential term $\E_0$, we consider the generalized Zeeman potential:
\beq
\E_0(\bm{n}) = m^2(1 - \bm{n}\cdot\bm{N})^p \ ,
\eeq
that is symmetric around the physical vacuum which is defined by 
\beq
\lim_{ r \to\infty} \bm{n}= \bm{N} \ ,
\label{finite-energy-condition3D}
\eeq
and $p$ is a positive real number defining the power of the potential. 
The latter condition~\eqref{finite-energy-condition3D} allows for the topological compactification of
the base space $\mathbb{R}^3$ into $S^3$, such that the vector field
becomes a map $\bm{n}: S^3 \to S^3$. This leads to a nontrivial
associated homotopy group $\pi_3(S^3) = \mathbb{Z}\ni B$, with $B$
being the topological degree or so-called ``baryon number''.
The topological degree can be computed as
\beq
B = \int\mathcal{B}\;\d^3x
= \frac{1}{2\pi^2}\int \epsilon_{abcd}n^a\p_1 n^b\p_2 n^c \p_3 n^d\;\d^3x,
\label{eq:topocharge}
\eeq
which is the pullback of the normalized volume form on the target
$S^3$ by the field $\bm{n}$.
Without loss of generality, as we will see in the next section, the vacuum can be chosen as
$\bm{N}=(0,0,0,1)^{\rm T}$ and in any case the potential breaks the symmetry from $\Og(4)$ to $\Og(3)$.
Finally, $E_1$ is the \emph{new} energy term, which we include for the first
time here, i.e.~the 3-dimensional generalization of the planar DMI, which we shall discuss in the next section. 

\subsection{Generalized DM Interaction}\label{sec:genDM}

\subsubsection{Bloch-type DMI}\label{sec:bloch-typeDM}

Since we want to write down a generalization of the 2-dimensional DM term of Bloch-type,
\beq
\kappa\epsilon^{iab}\p_in^a n^b, \qquad i=1,2,\qquad a,b=1,2,3,
\label{eq:DM2d_Bloch}
\eeq
for an $\SO(4)$ vector, we choose the $\SO(4)$ invariant tensor
$\epsilon^{abcd}$, which is totally antisymmetric and hence picks up a
sign flip under cyclic permutations: $\epsilon^{abcd}=-\epsilon^{dabc}$.
Now, in order to contract the $\epsilon$ tensor with the magnetization
vector, we notice that the antisymmetric nature of the tensor allows
us to contract only with one $n$.
As for derivatives of $n$, we can contract also only with one $\p n$,
since we wish to avoid Derrick's theorem by having less derivatives in
the DM term than in the kinetic term (i.e.~$(\p n)^2$).
Since the DM term breaks rotational invariance already in the standard 2D
formulation, there is no problem with contracting with a constant
tensor, $\Theta$.
We thus arrive at
\beq
\E^{\rm Bloch}_1 := \kappa \epsilon_{a b c d} \Theta^{abi} \p_i n^c n^d,
\label{eq:DM}
\eeq
where we raise and lower indices with Euclidean metrics.
Clearly, $\Theta$ is a mixed tensor having two $\SO(4)$-indices ($a$
and $b$) and one spatial index $i$; this may break rotational
invariance. 

In order to put the higher-dimensional DM term on the most general
standard form, we use $\SO(4)$ and $\SO(3)_s$ transformations to
simplify the tensor $\Theta$ as much as possible.
Since $\Theta$ is contracted with the $\SO(4)$-invariant tensor
$\epsilon_{abcd}$, we can from here on assume that it is antisymmetric
in the $\SO(4)$-indices $a$ and $b$:
\beq
\Theta^{abi} = -\Theta^{bai}.
\eeq
Performing simultaneously both orthogonal transformations,
$R\in\SO(4)$ and $r\in\SO(3)_s$, we get
\begin{align}
\E^{\rm Bloch}_1 &\to \kappa\epsilon_{abc'd'}\Theta^{abi'}r_{i'}^{\phantom{i'}i}\p_{i} (R^{c'}_{\phantom{c'}c}n^c) R^{d'}_{\phantom{d'}d}n^d\non
&= \kappa\epsilon_{a'b'c'd'}R^{a'}_{\phantom{a'}e}R_{a}^{\phantom{a}e}R^{b'}_{\phantom{b'}f}R_{b}^{\phantom{b}f}\Theta^{abi'}r_{i'}^{\phantom{i'}i}\p_{i} (R^{c'}_{\phantom{c'}c}n^c) R^{d'}_{\phantom{d'}d}n^d\non
&= \kappa\det(R)\epsilon_{abcd}R_{e}^{\phantom{e}a}R_{f}^{\phantom{f}b}\Theta^{efi'}r_{i'}^{\phantom{i'}i}\p_{i}n^c n^d,
\end{align}
from which we can read off the transformation on the tensor, $\Theta$,
as
\beq
\Theta^{abi}
\to R_{e}^{\phantom{e}a}R_{f}^{\phantom{f}b}\Theta^{efi'}r_{i'}^{\phantom{i'}i}.
\label{eq:Theta_transformation}
\eeq
Writing this in matrix notation for the $\SO(4)$-indices, we thus have
a simple orthogonal transformation:
\beq
\Theta^i \to R^{\rm T}\Theta^{i'}R \ r_{i'}^{\phantom{i'}i}.
\eeq
Now since $\Theta^i$ is an antisymmetric real $4\times4$ matrix for each
$i$, it has two conjugate pairs of pure imaginary eigenvalues.
There is, however, no reason for the matrices to be aligned \emph{a priori}.
This means that we can only diagonalize one of the three matrices,
say the $i=3$ matrix:
\beq
\Theta^3 \sim \diag(\i A^3,-\i A^3,\i F^3,-\i F^3),
\label{eq:Theta3_eigenvalues}
\eeq
with $\sim$ denoting under an orthogonal transformation.
Clearly, the determinant of $\Theta^3$ is positive as it must be.

Now we should recall that the kinetic term $\E_2$ is invariant under the full
$R\in\SO(4)$ transformation, but the vacuum, and consequently the potential $\E_0$ is not.
Notice that $\bm{N}$ will naturally transform under the rotation $R$.
We may utilize the $\SO(4)$ rotation to set the vacuum to be
\beq
\bm{N} = (0,0,0,1)^{\rm T},
\eeq
which is invariant only under the $\SO(3)_t$ subgroup represented by
$\rho\in\SO(3)_t\subset\SO(4)$ such that
\beq
R = \begin{pmatrix}
  \rho & 0\\
  0 & 1
\end{pmatrix}.
\label{eq:SO3subgroup}
\eeq
Parametrizing the tensor $\Theta^{abi}$ as 3 matrices that
transform under $\SO(4)$, we have
\beq
  \Theta^i =
  \frac12
\begin{pmatrix}
  0 & A^i & B^i & D^i\\
  -A^i & 0 & C^i & E^i\\
  -B^i & -C^i & 0 & F^i\\
  -D^i & -E^i & -F^i & 0
\end{pmatrix},
\label{eq:Theta_general_param}
\eeq
the rotation $\rho$ of Eq.~\eqref{eq:SO3subgroup} can only be used to
diagonalize the submatrix consisting of $A^i$, $B^i$ and $C^i$.
Since the submatrix is $3\times3$, it only possesses 1 purely imaginary
pair of nonvanishing eigenvalues, as well as a zero eigenvalue.
There is still no reason for the $\SO(3)_s$ vectors to be aligned, so
we can only diagonalize one of the $\Theta^i$ submatrices, say the $i=3$
again.
We thus arrive at
\beq
\Theta^{\mu} =
  \frac12
\begin{pmatrix}
  0 & A^{\mu} & B^{\mu} & D^{\mu}\\
  -A^{\mu} & 0 & C^{\mu} & E^{\mu}\\
  -B^{\mu} & -C^{\mu} & 0 & F^{\mu}\\
  -D^{\mu} & -E^{\mu} & -F^{\mu} & 0
\end{pmatrix}, \qquad
\Theta^3 =
\frac12
\begin{pmatrix}
  0 & A^3 & 0 & D^3\\
  -A^3 & 0 & 0 & E^3\\
  0 & 0 & 0 & F^3\\
  -D^3 & -E^3 & -F^3 & 0
\end{pmatrix},
\eeq
with $\mu=1,2$.
Using the $r\in\SO(3)_s$ degrees of freedom, we can eliminate one symbol
in $\Theta^{\mu}$, by rotating it into the $i=3$ direction; we may
choose this to be $F^{\mu}=0$.
We will denote this the standard form of the $\Theta$ tensor.
One result of this computation is that the minimal degrees of freedom
of the $\Theta$ tensor are 14.

At this point, it will be useful to identify a subset of the $\Theta$
tensor in standard form, that is invariant under a combined
$\SO(3)_t\subset\SO(4)$ and $\SO(3)_s$ symmetry.
The transformation is again that of
Eq.~\eqref{eq:Theta_transformation} with $R$ of
Eq.~\eqref{eq:SO3subgroup}, it takes the form
\beq
\Theta^i \to \frac12
\begin{pmatrix}
  \rho^{\rm T}\begin{pmatrix} 0 & A^j & B^j\\ -A^j & 0 & C^j\\ -B^j & -C^j & 0\end{pmatrix}\rho
    & \rho^{\rm T}\begin{pmatrix} D^j\\E^j\\F^j\end{pmatrix}\\
  \begin{pmatrix} -D^j & -E^j & -F^j\end{pmatrix}\rho & 0
\end{pmatrix} r_{j}^{\phantom{j}i},
\eeq
where $\rho\in\SO(3)_t$ and $r\in\SO(3)_s$.
From the above equation, it is clear that invariance requires the
triplet $A,B,C$ to transform under the adjoint representation of
$\SO(3)$ and the triplet $D,E,F$ to transform under the fundamental
representation.
This means that invariance under the combined rotation with the
locking of the two $\SO(3)$'s according to Eq.~\eqref{eq:SO3subgroup}
with $\rho=r$, is attained by the tensor
\beq
\Theta^i = \frac12
\begin{pmatrix}
  0 & \alpha\delta^{i3} & -\alpha\delta^{i2} & \beta\delta^{i1}\\
  -\alpha\delta^{i3} & 0 & \alpha\delta^{i1} & \beta\delta^{i2}\\
  \alpha\delta^{i2} & -\alpha\delta^{i1} & 0 & \beta\delta^{i3}\\
  -\beta\delta^{i1} & -\beta\delta^{i2} & -\beta\delta^{i3} & 0
\end{pmatrix},
\label{eq:Theta_inv_standard}
\eeq
with $\alpha,\beta\in\mathbb{R}$.
We see that the invariant tensor only has two degrees of freedom.
This invariant tensor can also be written in the basis of 't Hooft
tensors as
\beq
\Theta^i = \frac{\alpha+\beta}{4}\eta^i + \frac{\alpha-\beta}{4}\bar\eta^i,
\label{eq:Theta_inv_standard_tHooft}
\eeq
where the 't Hooft tensors are given by
\begin{align}
\eta^1 &=
\begin{pmatrix}
  0 & \tau^1\\
  -\tau^1 & 0
\end{pmatrix}, \qquad\quad\ 
\eta^2 =
\begin{pmatrix}
  0 & -\tau^3\\
  \tau^3 & 0
\end{pmatrix}, \qquad
\eta^3 =
\begin{pmatrix}
  \i\tau^2 & 0\\
  0 & \i\tau^2
\end{pmatrix},\non
\bar{\eta}^1 &=
\begin{pmatrix}
  0 & -\i\tau^2\\
  -\i\tau^2 & 0
\end{pmatrix}, \qquad
\bar{\eta}^2 =
\begin{pmatrix}
  0 & -\mathbf{1}_2\\
  \mathbf{1}_2 & 0
\end{pmatrix}, \qquad
\bar{\eta}^3 =
\begin{pmatrix}
  \i\tau^2 & 0\\
  0 & -\i\tau^2
\end{pmatrix}.
\end{align}

One may also consider the simplified tensor
\beq
\Theta^{abi} = \Gamma^a\delta^{bi},
\eeq
which clearly is not antisymmetric in the $\SO(4)$-indices $a$ and $b$.
Anti-symmetrizing them gives
\beq
\Theta^{abi} = \frac12\left(\Gamma^a\delta^{bi} - \Gamma^b\delta^{ai}\right),
\eeq
which written out in matrix form reads
\begin{align}
  \Theta^i = \frac12
  \begin{pmatrix}
  0 & \Gamma^1\delta^{2i}-\Gamma^2\delta^{1i} & \Gamma^1\delta^{3i}-\Gamma^3\delta^{1i} & -\Gamma^4\delta^{1i}\\
  \Gamma^2\delta^{1i}-\Gamma^1\delta^{2i} & 0 & \Gamma^2\delta^{3i}-\Gamma^3\delta^{2i} & -\Gamma^4\delta^{2i}\\
  \Gamma^3\delta^{1i}-\Gamma^1\delta^{3i} & \Gamma^3\delta^{2i}-\Gamma^2\delta^{3i} & 0 & -\Gamma^3\delta^{3i}\\
  \Gamma^4\delta^{1i} & \Gamma^4\delta^{2i} & \Gamma^4\delta^{3i} & 0
\end{pmatrix}.
\end{align}
This tensor is contained within the standard form of $\Theta$ by
identifying
\begin{align}
  A^i &= \left(-\Gamma^2,\Gamma^1,\sqrt{(\Gamma^1)^2+(\Gamma^2)^2}\right),\non
  B^{\mu} &= (-\Gamma^3,0),\non
  C^{\mu} &= (0,-\Gamma^3),\non
  D^{\mu} &= (-\Gamma^4,0),\non
  E^{\mu} &= (0,-\Gamma^4),\non
  F^3 &= -\Gamma^4,
\end{align}
and $F^{\mu}=0$ with $\mu=1,2$.
Considering an invariant subset of this tensor under the combined
$\SO(4)$ and $\SO(3)_s$ rotation \eqref{eq:Theta_transformation} yields
\beq
\Theta^{abi} \to \frac12(R^{\rm T}\Gamma)^a\delta^{bi} - \frac12(R^{\rm T}\Gamma)^b\delta^{ai}.
\eeq
Using now the rotation $R$ of Eq.~\eqref{eq:SO3subgroup}, it is clear
that invariance is only preserved if
\beq
\Gamma_{\rm inv}^a = -\beta\delta^{a4},
\label{eq:Gamma_inv}
\eeq
in agreement with the invariant tensor on the standard form
\eqref{eq:Theta_inv_standard}. 

Another simplified tensor may be constructed by taking the ``Hodge dual''
of the $\Theta$ tensor.
In order to dualize in 4 dimensions, we extend the index $i$ to
$i=1,2,3,4$, but with the derivative $\p_4=0$.
In this case, we have
\beq
\Theta^{abi} = \frac12\epsilon^{abic}\Psi^c,
\label{eq:Theta_Psi}
\eeq
where the factor of $1/2$ is introduced for convenience.
This tensor is also contained within the standard form of $\Theta$ by
identifying
\begin{align}
  A^3 &= \Psi^4,\non
  B^{\mu} &= (0,-\Psi^4),\non
  C^{\mu} &= (\Psi^4,0),\non
  D^{\mu} &= (0,\Psi^3),\non
  E^{\mu} &= (-\Psi^3,0),\non
  F^{\mu} &= (\Psi^2,-\Psi^1),
\end{align}
where this time we have chosen $A^{\mu}=0$.
It is again clear that only a subset of this tensor is an invariant
under the transformation \eqref{eq:Theta_transformation} and it is
given by
\beq
\Psi^a = \alpha\delta^{a4}.
\label{eq:Theta_Psi_inv}
\eeq
Interestingly, the invariant of this tensor corresponds to the
$\alpha$-part of the invariant tensor on standard form
\eqref{eq:Theta_inv_standard}, whereas the invariant tensor
\eqref{eq:Gamma_inv} corresponds to the $\beta$-part.

\subsubsection{N\'eel-type DMI}

In two dimensions the DM term can equivalently take a different form,
known as the DMI of N\'eel type:
\beq
\kappa(\bm{n}\cdot\nabla n^3 - n^3\nabla\cdot\bm{n}),
\label{eq:DM2d_Neel}
\eeq
with $\nabla=(\p_1,\p_2,0)$ and $\bm{n}=(n^1,n^2,n^3)$, which is due
to the Rashba spin-orbit coupling (SOC), whereas the Bloch-type DMI
\eqref{eq:DM2d_Bloch} corresponds to the Dresselhaus SOC.
Although, this DM term looks quite different from the Bloch type
\eqref{eq:DM2d_Bloch}, the N\'eel-type DMI is simply obtained from the
Bloch-type DMI by performing an $\SO(2)_t\subset\SO(3)_t$ rotation of the
magnetization vector $(n^1,n^2,n^3)\mapsto(n^2,-n^1,n^3)$ and keeping
the derivative vector fixed.
The generalization of the above 2-dimensional DM term to a
higher-dimensional DM term, however, is straightforward and somewhat
different\footnote{Often the opposite sign of the N\'eel-type DMI is
used, which can be reached from the Bloch-type DMI simply by rotating 
the opposite way in the 1-2 plane: $(n^1,n^2,n^3)\mapsto(-n^2,n^1,n^3)$,
or alternatively by flipping the sign of $\kappa$.}:
\beq
\E^{\text{N\'eel}}_1 = \kappa(n^4\nabla\cdot\bm{n} - \bm{n}\cdot\nabla n^4),
\label{eq:DM_Neel}
\eeq
where now $\nabla=(\p_1,\p_2,\p_3,0)$ and $\bm{n}=(n^1,n^2,n^3,n^4)$. 
The symmetry properties are also quite clear and simple in this case,
since the inner product is preserved by a combined rotation
in $\rho\in\SO(3)_t\subset\SO(4)$ (see Eq.~\eqref{eq:SO3subgroup}) and
$r\in\SO(3)_s$, with the symmetry-locking mechanism again given by $\rho=r$.

In fact, by a closer inspection it is clearly $\SO(3)_{\rm diag}$ invariant, as
it is simply equal to the tensor \eqref{eq:Theta_Psi} with $\Psi^a$ of 
Eq.~\eqref{eq:Theta_Psi_inv}, corresponding to the $\alpha$-part of
the $\SO(3)_{\rm diag}$ invariant tensor \eqref{eq:Theta_inv_standard}.
Indeed, a simple calculation shows that
\beq
\frac12\kappa\epsilon_{abcd}\epsilon^{abi4}\p_in^c n^d
=\kappa(\delta_c^i\delta_d^4-\delta_c^4\delta_d^i)\p_in^c n^d
=\kappa(n^4\nabla\cdot\bm{n} - \bm{n}\cdot\nabla n^4),
\eeq
which is exactly Eq.~\eqref{eq:DM_Neel}.

\subsubsection{Axially symmetric DMI}

The N\'eel-type DM term \eqref{eq:DM_Neel} already possesses
$\SO(3)_{\rm diag}$
symmetry, whereas the general Bloch-type generalization \eqref{eq:DM}
does not.
Considering the $\Theta$ tensor \eqref{eq:Theta_general_param}, we can
relax the spherical (i.e.~$\SO(3)$) symmetry restriction to only an
axial symmetry.
Using the transformation \eqref{eq:Theta_transformation}, but with the
$\SO(4)$ transformation, i.e.,
\beq
R = \begin{pmatrix}
  \rho' & 0\\
  0 & \mathbf{1}_2
\end{pmatrix},
\eeq
with $\rho'\in\SO(2)_t$, $\mathbf{1}_2$ being the $2\times2$ unit matrix,
and with the $\SO(2)_s\subset\SO(3)_s$ spatial rotation
\beq
r = \begin{pmatrix}
  \rho' & 0\\
  0 & 1
\end{pmatrix},
\eeq
we can find an axially invariant $\Theta$ tensor of the form
\beq
\Theta^i = \frac12
\begin{pmatrix}
  0 & 0 & \gamma\delta^{i1} & \eta\delta^{i1}\\
  0 & 0 & \gamma\delta^{i2} & \eta\delta^{i2}\\
  -\gamma\delta^{i1} & -\gamma\delta^{i2} & 0 & F^3\delta^{i3}\\
  -\eta\delta^{i1} & -\eta\delta^{i2} & -F^3\delta^{i3} & 0
\end{pmatrix}.
\eeq
The two vectors proportional to $\gamma$ and $\eta$ are
invariants under the combined $\SO(2)$ rotation in the $x$-$y$ plane,
but $(F^1,F^2)$ still transform, so we must set them to zero.

\subsection{Spherically symmetric magnetic Skyrmions}\label{sec:sph_DM_Sk}

In order to construct a spherically symmetric magnetic Skyrmion, we
necessarily need to consider only the $\SO(3)_{\rm diag}$ invariant subtensors,
which for the Bloch-type DM term corresponds to the $\alpha$ and
$\beta$ parts of Eq.~\eqref{eq:Theta_inv_standard}.
Considering the model that closest resembles the 2D magnetic Skyrmion,
we take the energy density to be
\beq
\E = \E_2 + \E_1 + \E_0.
\eeq
The kinetic term, $\E_2$, is invariant under
$\Og(4)_{t}\times\Og(3)_{s}\ltimes T_3$ symmetry, which
is broken down to $\Og(3)_{t}\times\Og(3)_{s}\ltimes T_3$
by the potential, $\E_0$, where $T_3$ are spatial translations and
$\ltimes$ is the semi-direct product.
The DM term, $\E_1$, further breaks the symmetry which is clear from
the derivation in Sec.~\ref{sec:bloch-typeDM}, viz.~the internal
symmetry is locked with the spatial rotations, yielding the symmetry
of the theory
\beq
G = \Og(3)_{\rm diag}\ltimes T_3.
\eeq
The hedgehog Ansatz with an azimuthal rotation by $\delta$ is 
given by the map
\beq
\bm{n} = \begin{pmatrix}
  \sin\chi\sin\theta\cos(\phi+\delta)\\
  \sin\chi\sin\theta\sin(\phi+\delta)\\
  \sin\chi\cos\theta\\
  \cos\chi
\end{pmatrix}, \qquad
\chi,\theta\in[0,\pi], \qquad
\varphi\in[0,2\pi),
\label{eq:hedgehog}
\eeq
with $(r,\theta,\phi)$ being the normal spherical coordinates and
$\chi(r)$ the radial profile function.
The energy densities for the kinetic and potential terms are independent of
$\delta$ and read
\begin{align}
  \E_2 &= \frac12(\chi')^2 + \frac{\sin^2\chi}{r^2},\\
  \E_0 &= m^2(1-\cos\chi)^p, \qquad p>0,
\end{align}
with $\chi':=\p_r\chi(r)$.
We typically take $p=1$ or alternatively $p=\frac32$ or $p=2$.

More interesting is the $\SO(3)_{\rm diag}$-invariant DM term
\eqref{eq:Theta_inv_standard}, which under the hedgehog Ansatz reduces
to
\begin{align}
  \E_1^{\rm Bloch} &= 
  \kappa\alpha\cos\delta\left(\sin^2\theta\chi' + \frac{(1+\cos^2\theta)\sin2\chi}{2r}\right)
  + \kappa\alpha\left(\cos^2\theta\chi' + \frac{\sin^2\theta\sin2\chi}{2r}\right)\non
  &\phantom{=\ }
  + 2\kappa\beta\sin\delta\,\frac{\cos\theta\sin^2\chi}{r}.
  \label{eq:block-type-DMI-spheric}
\end{align}
The $\beta$-part of the term vanishes upon integration of $\d\theta$
(i.e.~$\int_0^\pi\sin2\theta\d\theta=0$).
We notice that the term becomes spherically symmetric when $\delta=0$:
\beq
\E_1^{\rm Bloch} = \kappa\alpha\left(\chi' + \frac{\sin2\chi}{r}\right).
\eeq
Integrating instead~\eqref{eq:block-type-DMI-spheric} over $\theta$, we obtain
\beq
\int \E_1^{\rm Bloch}\sin\theta\d\theta
= \frac{2\kappa\alpha}{3}(1+2\cos\delta)\left(\chi' + \frac{\sin2\chi}{r}\right).
\eeq
Since the DM term contributes as negative energy, maximizing its
prefactor corresponds to minimizing the energy, which is quickly seen
to occur at $\delta=0$, the symmetric point.
We also absorb $\alpha$ into $\kappa$ and thus arrive at
\beq
\E_1^{\rm Bloch} = \kappa\left(\chi' + \frac{\sin2\chi}{r}\right).
\label{eq:DM_hedgehog}
\eeq

If we instead consider the generalized N\'eel-type DM term
\eqref{eq:DM_Neel}, we obtain under the hedgehog Ansatz  
exactly the $\alpha$-part of the Bloch-type DM (see
Eq.~\eqref{eq:block-type-DMI-spheric}), as expected.
The symmetry is enhanced when $\delta=0$ and in that case the above
density reduces to
\beq
\E_1^{\text{N\'eel}} = \kappa\left(\chi' + \frac{\sin2\chi}{r}\right).
\label{eq:E1-neel}
\eeq
We thus conclude that both the Bloch-type and N\'eel-type DMIs give
under the hedgehog Ansatz \eqref{eq:hedgehog} rise to the same DM
term, i.e.~$E_1^{\text{Bloch}}=E_1^{\text{N\'eel}} \equiv E_1$, which is schematically very similar to the polar symmetric one in two dimensions. A summary is presented in Tab.~\ref{tab:DMIs} for the reader's convenience.

\begin{table}[!htp]
  \centering
  \begin{tabular}{c||ccc}
    & $\alpha$-part  & $\beta$-part & sph.~sym. \\
    \hline\hline
    Bloch-type & \eqref{eq:Theta_Psi_inv} & \eqref{eq:Gamma_inv} & \eqref{eq:DM_hedgehog}~($\beta=0$) \\
    N\'eel-type & \eqref{eq:Theta_Psi_inv} & 0 & \eqref{eq:E1-neel} \\
  \end{tabular}
  \caption{Bloch-type and N\'eel-type DMIs. Note that the $\beta$-part
      of the Bloch-type DM vanishes in the spherically symmetric case.
    }
  \label{tab:DMIs}
\end{table}

The total energy ($E=\int\E\d^3x$) of the higher-dimensional magnetic
Skyrme model under the hedgehog Ansatz \eqref{eq:hedgehog} is thus
given by 
\beq
E = 4\pi\int\left[
  \frac12r^2(\chi')^2
  +\sin^2\chi
  +\kappa(r^2\chi' + r\sin2\chi)
  +m^2 r^2(1-\cos\chi)^p
  \right] \d r,
\eeq
which gives rise to the following Euler-Lagrange equation
\beq
\eom_\chi = \chi'' + \frac2r\chi' - \frac{\sin2\chi}{r^2}
+\frac{4\kappa\sin^2\chi}{r}
-p m^2(1-\cos\chi)^{p-1}\sin\chi = 0.
\label{eq:eom_chi}
\eeq

It is well known that using a scaling argument, we can without loss of
generality reduce the parameters of the theory from 2 to 1, by
rescaling lengths and choosing a suitable energy unit.
Rescaling $r\to\lambda r$, we thus arrive at
\beq
E_\lambda = \lambda E_2 + \lambda^2 E_1 + \lambda^3 E_0.
\label{eq:Elambda}
\eeq
Derrick's theorem \cite{Derrick:1964ww} prevents stable solitonic
solutions to this theory if all $E_{2,1,0}$ were positive definite.
However, since $E_1$ can be negative, stability can be attained.
From this argument, it is clear that we could remove the potential
term $E_0$ and, in principle, still have stable solutions.
We therefore scale away $\kappa$ with $\lambda=\kappa^{-1}$ and
energies are measured in units of $\lambda=\kappa^{-1}$.
Absorbing $\lambda$ into the mass parameter, we arrive at
\beq
E = \frac{4\pi}{\kappa}\int\left[
  \frac12r^2(\chi')^2
  +\sin^2\chi
  +(r^2\chi' + r\sin2\chi)
  +m^2 r^2(1-\cos\chi)^p
  \right] \d r.
\label{eq:E_rescaled}
\eeq
Since $\kappa=0$ cannot give rise to stable solitons, it is no problem
to divide the energy functional by $\kappa$.
The equation of motion is just Eq.~\eqref{eq:eom_chi} with $\kappa$
set to unity.

Finally, the topological charge is given by Eq.~\eqref{eq:topocharge},
which under the Ansatz \eqref{eq:hedgehog} simplifies to
\beq
B = \int\mathcal{B}\;\d^3x
= -\frac{2}{\pi} \int \sin^2(\chi) \chi'\; \d r
= \frac{2}{\pi}\int_0^\pi \sin^2\chi\;\d\chi = 1,
\eeq
where we have used the boundary conditions $\chi(0)=\pi$ and
$\chi(\infty)=0$.

\subsubsection{Full equations of motion}

Using the principle of symmetric criticality, the equation of
motion~\eqref{eq:eom_chi} solves the full equations of motion, if and
only if the Ansatz is compatible with the symmetries of the full
theory.
Nevertheless, the full equations of motion serve as a consistency
check of the Ansatz and in the 2D case, they furthermore determine the
phase parameter $\delta$.

The full equations of motion read
\beq
\p_i^2n^a
+ 2\kappa\epsilon_{abcd}\Theta^{bci}\p_i n^d
+ p m^2(1-\bm{n}\cdot\bm{N})^{p-1}N^a
- \lambda n^a = 0,
\label{eq:eom_full}
\eeq
where $\lambda$ is a Lagrange multiplier that can be determined as
\beq
\lambda = \bm{n}\cdot\p_i^2\bm{n}
- 2\kappa\epsilon_{abcd}\Theta^{abi}\p_i n^c n^d
+ p m^2(1-\bm{n}\cdot\bm{N})^{p-1}\bm{n}\cdot\bm{N}.
\eeq
Inserting the spherical Ansatz~\eqref{eq:hedgehog}, Eq.~\eqref{eq:eom_full} reduces to
\beq
\frac{\p\bm{n}}{\p\chi}\times\eom_\chi
+ \sin\left(\frac\delta2\right)\bm{V} + \sin^2\left(\frac\delta2\right)\bm{W} = 0,
\eeq
with $\eom_\chi$ of Eq.~\eqref{eq:eom_chi} and where we have defined
\begin{align}
  \bm{V} &= 2\kappa\begin{pmatrix}
    -\frac12\sin\chi\sin\theta\left[3\sin\left(\frac\delta2+\phi\right)+\sin\left(\frac{3\delta}{2}+\phi\right)\right]\chi'\\
    \phantom{+}
    \frac12\sin\chi\sin\theta\left[3\cos\left(\frac\delta2+\phi\right)+\cos\left(\frac{3\delta}{2}+\phi\right)\right]\chi'\\
    0\\0
  \end{pmatrix},\\
  \bm{W} &= 2\kappa\begin{pmatrix}
    n^1\left[-(1+\cos^2\theta)\frac{\sin2\chi}{r}+\cos(2\theta)\chi'\right]\\
    n^2\left[-(1+\cos^2\theta)\frac{\sin2\chi}{r}+\cos(2\theta)\chi'\right]\\
    -n^3\left[(1+\cos^2\theta)\frac{\sin2\chi}{r}+2\sin^2(\theta)\chi'\right]\\
    2(1+\cos^2\theta)\frac{\sin^3\chi}{r}
  \end{pmatrix},
\end{align}
which demonstrates that Eq.~\eqref{eq:eom_chi} is consistently solving
the full equations of motion if and only if $\delta=0$.
Unlike the situation in 2D, here the solution to the full equations of
motion uniquely determine the rotation phase of the Ansatz to be
$\delta=0$.
On the other hand, in the standard 2D magnetic Skyrmion case, there
are two solutions corresponding to a rotation in the plane by
$\delta=\tfrac\pi2$ and $\delta=-\tfrac\pi2$, with the former being
the minimum of the energy and the latter the maximum.
In this case, the generalized DMI is only spherically symmetric when
$\delta=0$ and therefore $\eom_\chi=0$ only solves the full equations
of motion for that value of $\delta$.

\subsubsection{Numerical Skyrmion solutions}

We will now compute a spectrum of numerical solutions for the 3D
generalization of the Skyrmion with a higher-dimensional DM term that is
$\SO(3)_{\rm diag}$-invariant: namely the $\alpha$-invariant part of
Eq.~\eqref{eq:Theta_inv_standard}.
We utilize the rescaling of lengths, so that the coefficient of the
higher-dimensional DMI is unity, but the energy is measured in units of
inverse $\kappa$, see Eq.~\eqref{eq:E_rescaled}.
We solve the equation of motion \eqref{eq:eom_chi} with $\kappa$ set
to unity (under the rescaled scheme) using gradient flow.
The gradient flow method is suitable for the stable 3D Skyrmions and
is easier to use than a shooting method due to the large sizes of the
Skyrmions, when $m$ is small. 
We have cross-checked the solutions using the shooting algorithm 
and confirmed their accuracy.

\begin{figure}[!htp]
  \centering
  \mbox{\subfloat[$p=1$]{\includegraphics[width=0.4\linewidth]{{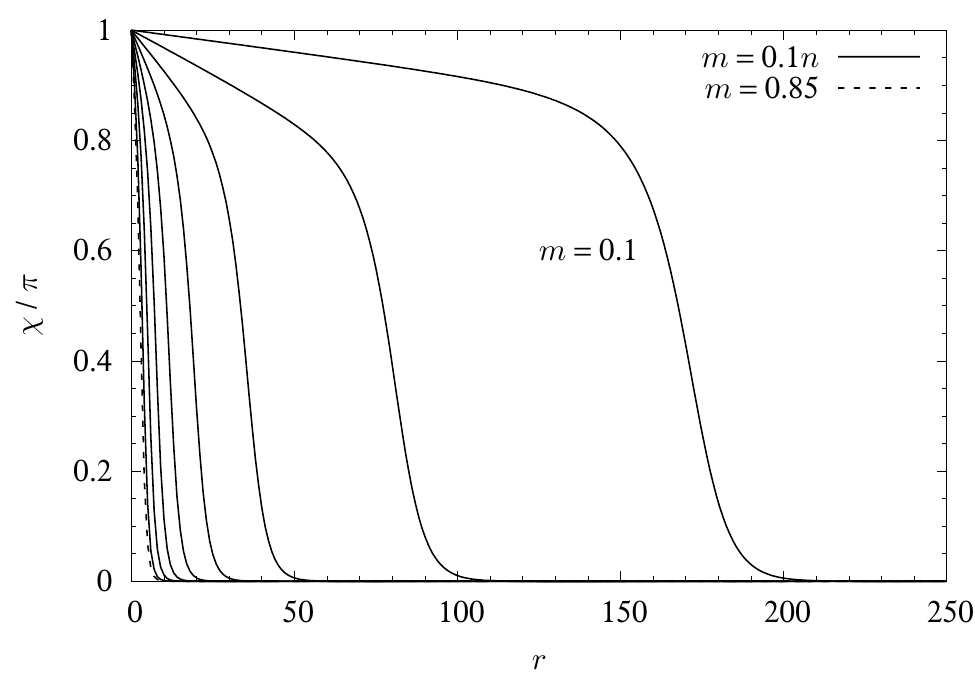}}}
    \subfloat[$p=1$]{\includegraphics[width=0.4\linewidth]{{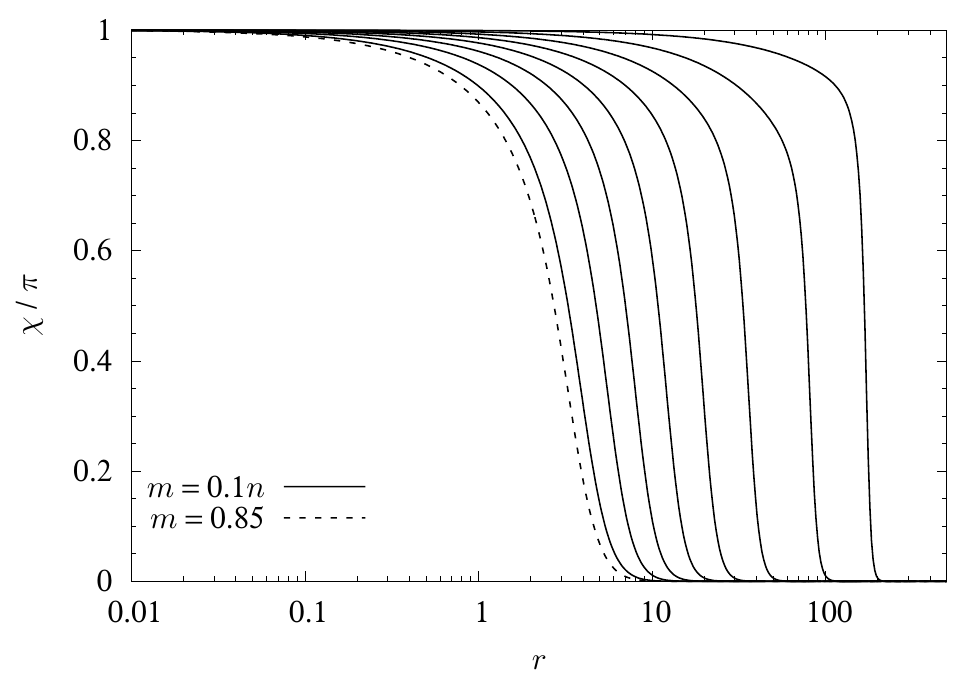}}}}
  \mbox{\subfloat[$p=\frac32$]{\includegraphics[width=0.4\linewidth]{{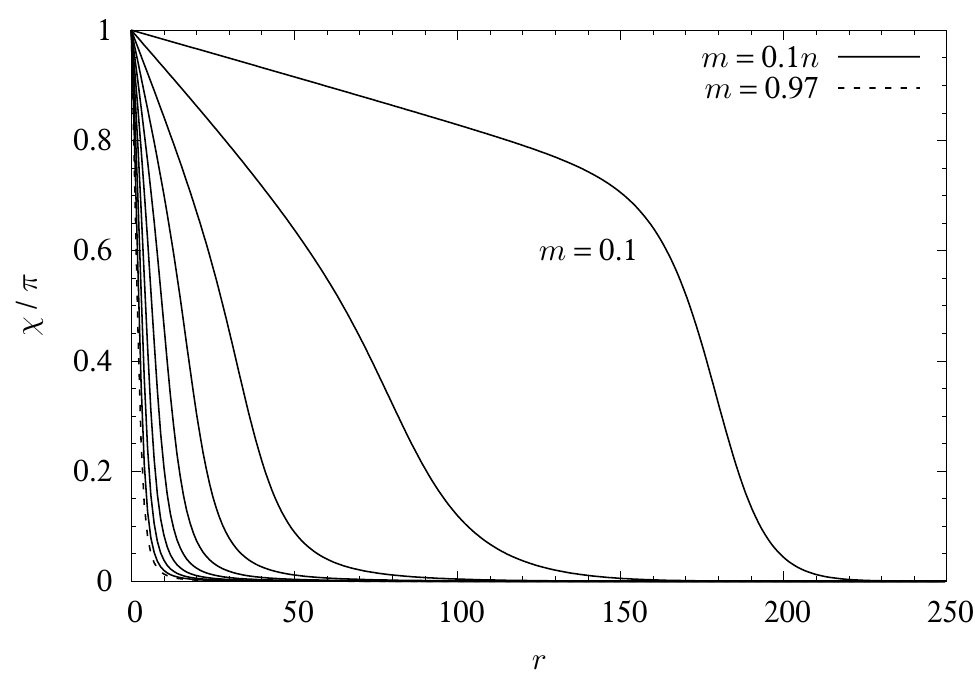}}}
    \subfloat[$p=\frac32$]{\includegraphics[width=0.4\linewidth]{{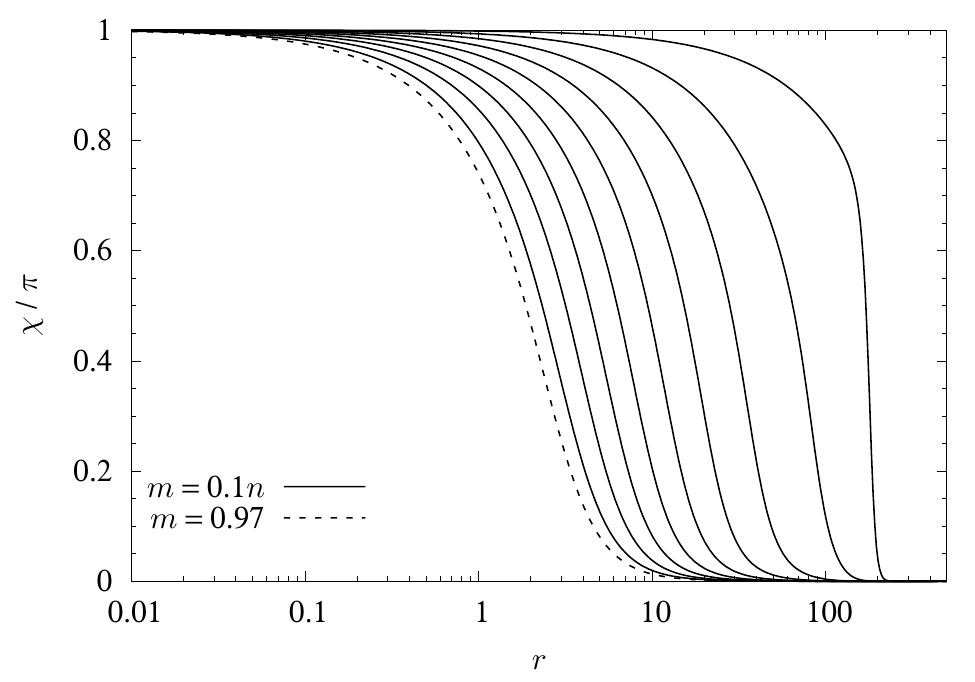}}}}
  \mbox{\subfloat[$p=2$]{\includegraphics[width=0.4\linewidth]{{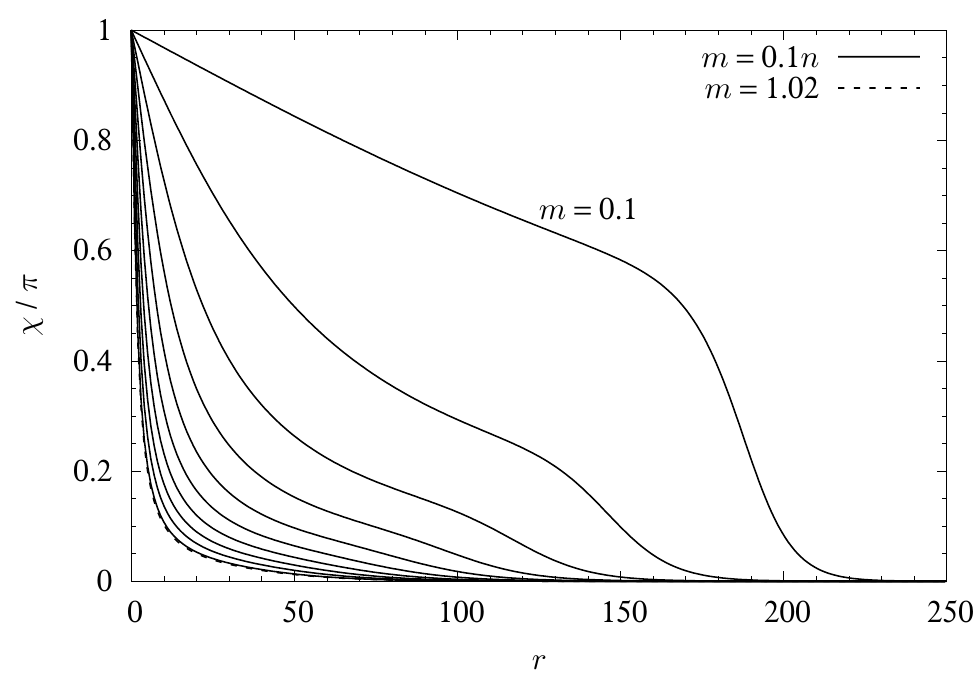}}}
    \subfloat[$p=2$]{\includegraphics[width=0.4\linewidth]{{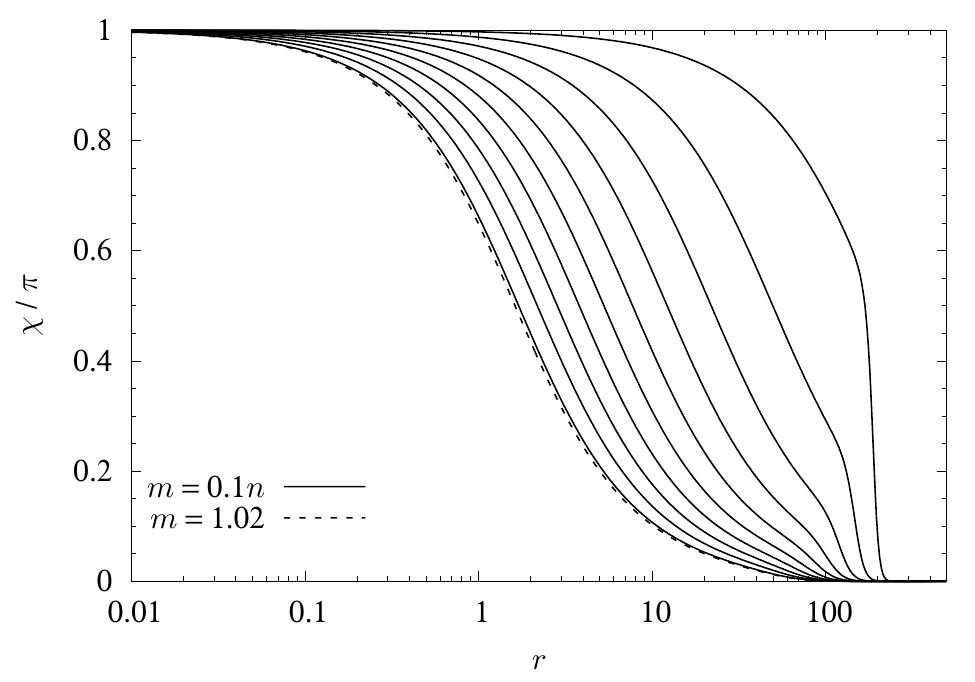}}}}
  \caption{Profile functions for 3D magnetic Skyrmions with higher-dimensional
    $\SO(3)_{\rm diag}$-invariant DM term. Left panels (a,c,e) show the profile
    on a regular scale and the right panels (b,d,f) on a log scale.
    The $m_{\rm crit}$ solution is shown with a dashed line.
    The rows of panels correspond to the potential power parameter
    $p=1,\tfrac32,2$. No solutions exist for $m>m_{\rm crit}$
    nor for $m=0$.
  }
  \label{fig:sols}
\end{figure}
\begin{table}[!htp]
  \centering
  \begin{tabular}{c||c}
    $p$ & $m_{\rm crit}$\\
    \hline\hline
    1 & 0.85\\
    $\tfrac32$ & 0.97\\
    2 & 1.02
  \end{tabular}
  \caption{The critical mass parameter $m_{\rm crit}$ for the
    potential with $p=1,\tfrac32,2$. No solutions exist for
    $m>m_{\rm crit}$, but solutions exist in the interval
    $m\in(0,m_{\rm crit}]$.
  }
  \label{tab:mcrit}
\end{table}
Fig.~\ref{fig:sols} shows the profile functions of the solutions for
the full range of parameters, i.e.~$m\in(0,m_{\rm crit}]$ in the
rescaled lengths where $\kappa$ only enters as the (inverse) unit of
energy.
There exist no solution for $m>m_{\rm crit}$ and for the 3
potential power parameters, $p=1,\tfrac32,2$, we give the approximate
critical masses in Tab.~\ref{tab:mcrit}.
Notice also that no 3D magnetic Skyrmion solution exists for $m=0$,
see below though.
We can understand the critical mass as follows.
For fixed DM-term coefficient (in the effective rescaling of
parameters in the theory), the only theory parameter is $m$.
For small $m$, the DM term can stabilize the soliton, whereas for
large $m$, the pressure to collapse the soliton eventually becomes too
large, and hence no soliton exists.

As can be inferred from Derrick scaling, the largest 3D magnetic
Skyrmion solution is the one with the smallest mass and the smallest
ones are the $m=m_{\rm crit}$ ones, see the approximate sizes in
Fig.~\ref{fig:size}(a).

\begin{figure}[!htp]
  \centering
  \mbox{\subfloat[$p=1$]{\includegraphics[width=0.4\linewidth]{{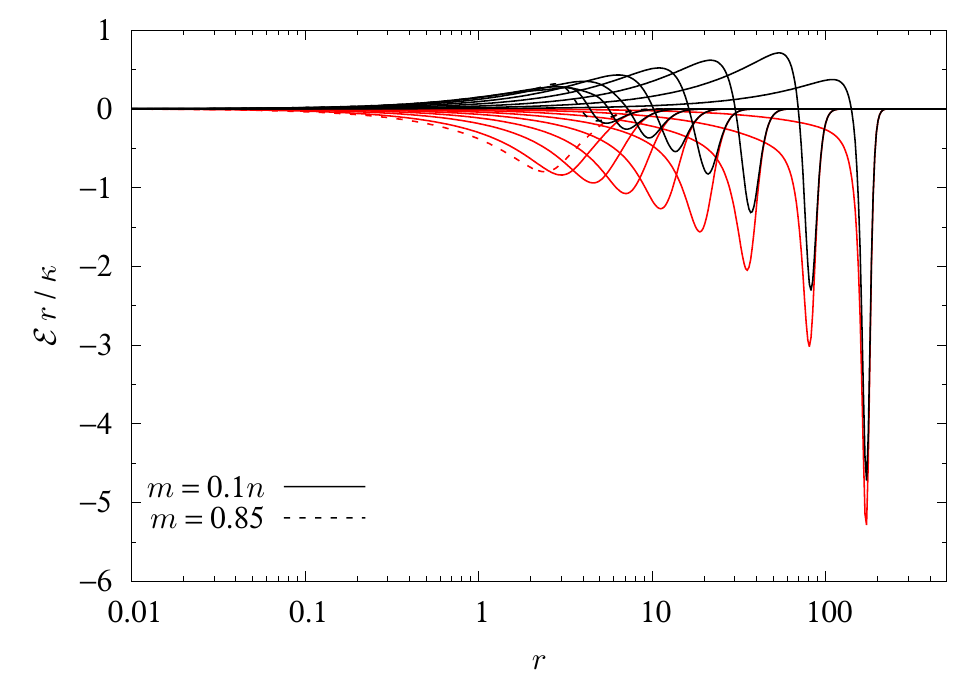}}}
    \subfloat[$p=1$]{\includegraphics[width=0.4\linewidth]{{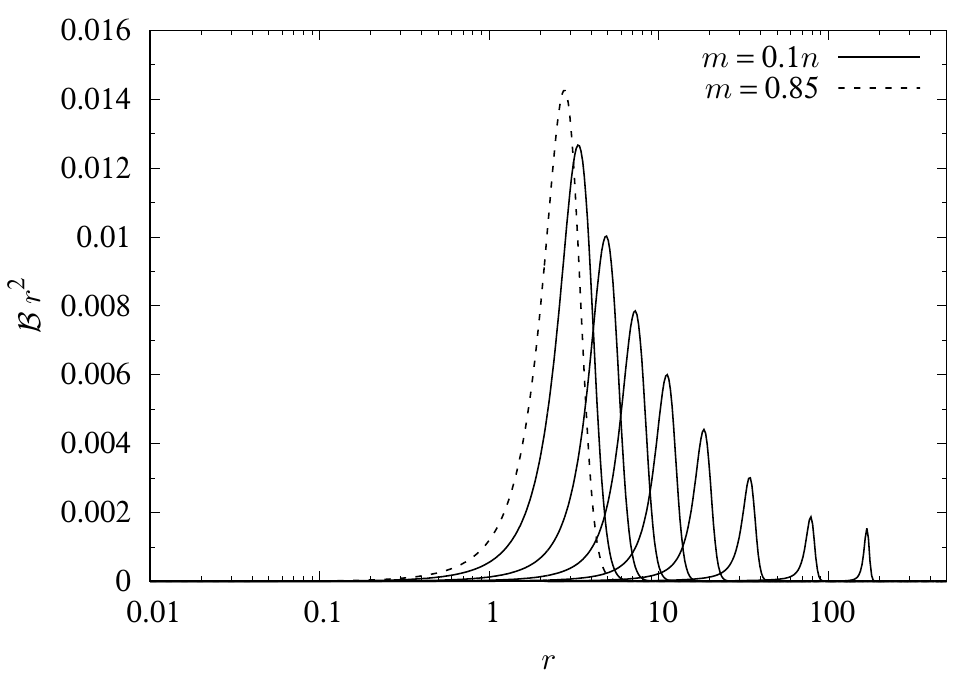}}}}
  \mbox{\subfloat[$p=\frac32$]{\includegraphics[width=0.4\linewidth]{{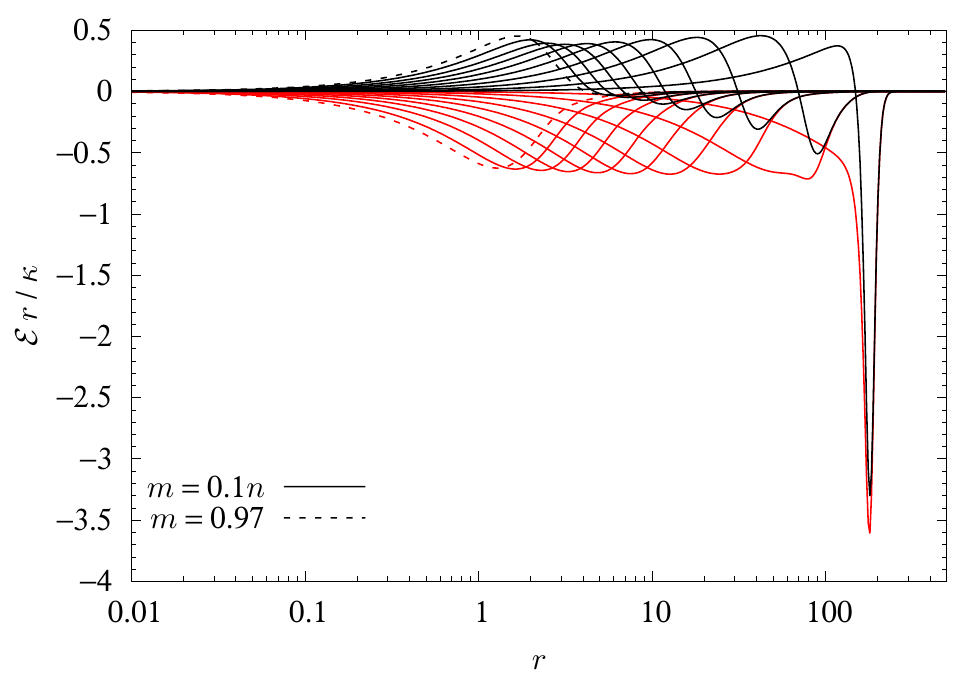}}}
    \subfloat[$p=\frac32$]{\includegraphics[width=0.4\linewidth]{{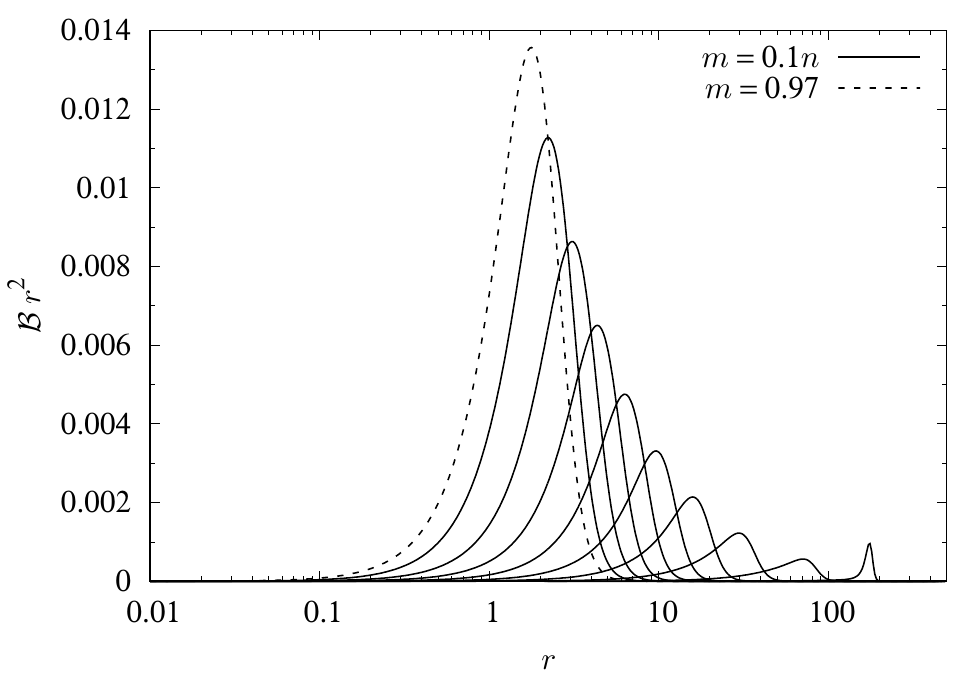}}}}
  \mbox{\subfloat[$p=2$]{\includegraphics[width=0.4\linewidth]{{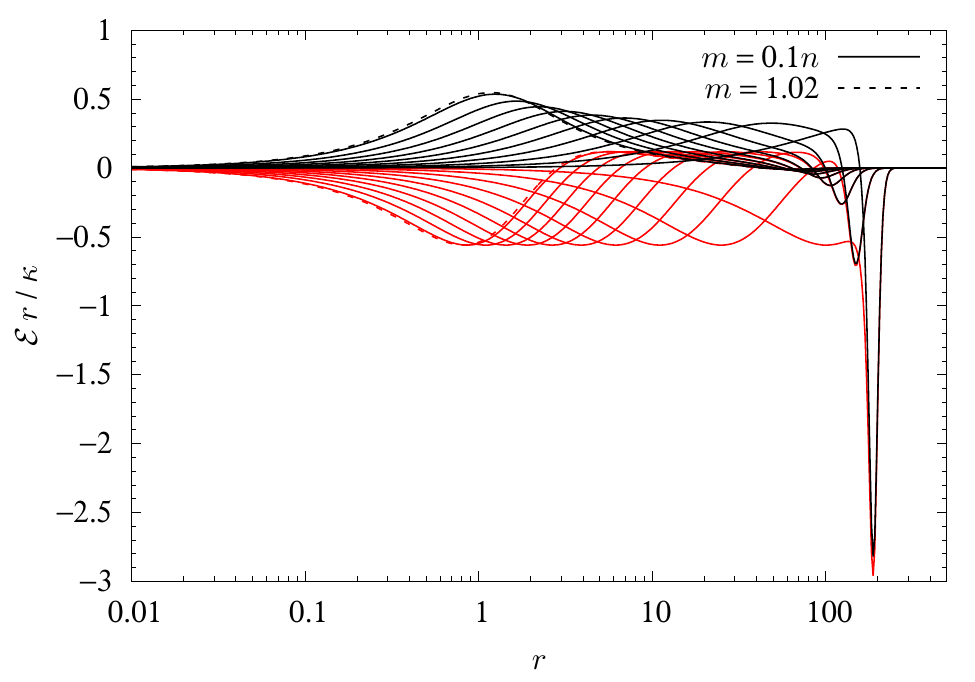}}}
    \subfloat[$p=2$]{\includegraphics[width=0.4\linewidth]{{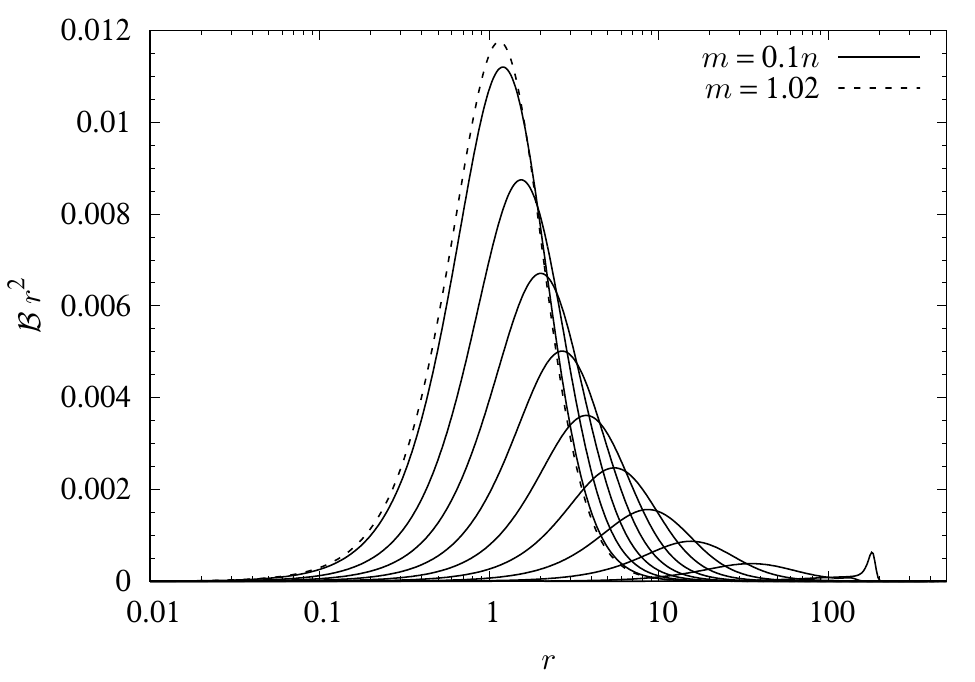}}}}
  \caption{Energy density and topological charge density for 3D magnetic
    Skyrmions with higher-dimensional $\SO(3)_{\rm diag}$-invariant DM term.
  Left panels (a,c,e) show the total energy densities weighted with
  $r$ (for visualization purposes) as black lines and the DM
  contribution as red lines. Right panels (b,d,f) show the topological charge density, weighted
  with $r^2$.
  The $m_{\rm crit}$ solution is shown with a dashed line.
  The rows of panels correspond to the potential power parameter
  $p=1,\tfrac32,2$. No solutions exist for $m>m_{\rm crit}$
    nor for $m=0$.
  }
  \label{fig:ensbs}
\end{figure}
Fig.~\ref{fig:ensbs} shows the energy densities (left panels) and
topological charge densities (right panels).
The left panels show the total energy densities for the full range of
the mass parameter, $m$, using black lines and the DM contribution to
the energy density using red lines. 
The densities are weighted by $r^2$ to make the features of the
densities better visible.
The 3 rows of the figure correspond to different values of the
potential power parameter, $p=1,\tfrac32,2$.

\subsubsection{Restricted model}

If we consider a rescaling of lengths $r\to\lambda r$ with
$\lambda=\frac{\kappa}{m^2}$, we obtain a theory where the coefficients
in front of the DM term and the potential term both equal
$\frac{\kappa^2}{m^2}$, whereas the coefficient of the kinetic term
remains unity (the energy unit is now $\frac{\kappa}{m^2}$).
Now, taking $\kappa\gg m>0$, the kinetic term becomes negligible
and we arrive at the restricted version of the model,
\begin{align}
  E^{\rm restricted} &= \frac{\kappa^2}{m^4}E_1 + \frac{\kappa^3}{m^6}E_0\non
  &= \frac{4\pi\kappa^3}{m^4}\int\left[
    r^2\chi' + r\sin2\chi + r^2(1-\cos\chi)^p\right]\d r,
\end{align}
which has the algebraic and explicit equation of motion
\beq
r = \frac{4\sin\chi}{2^{p-1}p\sin^{2(p-1)}\left(\frac\chi2\right)}.
\eeq
This solution is formally identical with the restricted 2D magnetic
Skyrmion solution of Ref.~\cite{Bolognesi:2024mjs}.
In particular, we have 
\begin{align}
  p &= 1\;: \qquad
  \chi(r) = \pi - \arcsin\left(\frac{r}{4}\right), \\
  p &= \tfrac32\;: \qquad
  \chi(r) = 2\arccos\left(\frac{3\sqrt{2}r}{16}\right), \\
  p &= 2\;: \qquad
  \chi(r) = 2\arctan\left(\frac{r}{2}\right).
\end{align}

\subsubsection{The anti-Skyrmion}
In the 2-dimensional magnetic Skyrme model, the anti-Skyrmion is
unstable.
It is worthwhile to discuss the anti-Skyrmion in this
higher-dimensional model, since it is somewhat different from the
2-dimensional case.
The simplest way to obtain the anti-Skyrmion is to flip the polar
angle $\phi\to-\phi$ in Eq.~\eqref{eq:hedgehog}.
The $\beta$-part of the DMI still vanishes, but the $\alpha$-part
changes to
\beq
\int\E_1^{\overline{\rm Sk}}\sin\theta\;\d\theta = \frac{2\kappa}{3}\left[
  (1-\cos\delta)\frac{\sin2\chi}{r} + (1+2\cos\delta)\chi'
  \right].
\eeq
Integrating by parts and assuming $m>0$ such that the boundary term
vanishes, we can show that
\beq
\int\E_1^{\overline{\rm Sk}}\sin\theta\;\d\theta = \frac{2\kappa}{3}\left[
  \cos\delta(3-\sin^2\chi) + 2\sin^2\chi\right]\chi',
\eeq
which should be compared to
\beq
\int\E_1^{\rm Sk}\sin\theta\;\d\theta = \frac{2\kappa}{3}\left[(1+2\cos\delta)2\sin^2(\chi)\right]\chi',
\eeq
for the Skyrmion.
For the Skyrmion the prefactor is clearly maximized by $\delta=0$.
This is also the case for the anti-Skyrmion, which can be seen by
comparing the case with $\kappa>0$ and $\delta=0$ to the case of
$\kappa<0$ and $\delta=\pi$, which are the most positive and the
most negative values of the square brackets.
So although the anti-Skyrmion exists, it will have a smaller
coefficient of the DM term and hence a larger energy than the
Skyrmion for $\delta:=0$.
This can be seen by comparing the DMI for the anti-Skyrmion to the
Skyrmion in the following form:
\begin{align}
  \int\E_1^{\rm Sk}\sin\theta\;\d\theta &= 2\kappa\left(\frac{\sin2\chi}{r}+\chi'\right),\\
  \int\E_1^{\overline{\rm Sk}}\sin\theta\;\d\theta &= \int\E_1^{\rm Sk}\sin\theta\;\d\theta
  - 2\kappa\frac{\sin2\chi}{r}.
\end{align}
Since the term $-\frac{\sin2\chi}{r}$ is positive for small $r$
where $\chi>\pi/2$, it will increase the energy of the anti-Skyrmion
with respect to the Skyrmion (more than it will decrease the energy
for $\chi>\pi/2$, since the term will be small for large $r$),
leaving the Skyrmion the stable soliton in the theory.

\subsection{Magnetic sphalerons}\label{sec:sphalerons}

Considering again the rescaling of lengths as $r\to\lambda r$, we get
Eq.~\eqref{eq:Elambda}. By identifying the characteristic radius of
the soliton as $R$ and approximating the integrals of the energy
functional by constants as
\beq
\lambda E_2 = c_2 R, \qquad
\lambda^2 E_1 = -c_1 \kappa R^2, \qquad
\lambda^3 E_0 = c_0 m^2 R^3,
\eeq
we can write the approximate energy functional as a function of the
soliton size, $R$:
\beq
E(R) = c_2 R - c_1\kappa R^2 + c_0 m^2 R^3,
\label{eq:ER}
\eeq
Choosing energy and length units more convenient for the problem at
hand, we can write the above function as
\beq
E(\tR) = \frac{c_2^{\frac32}}{\sqrt{c_0}m}\left[\tR - \tilde\kappa\tR^2 + \tR^3\right],\qquad
\tR = \frac{R}{R_0} = \sqrt{\frac{c_0}{c_2}} m R,
\label{eq:Rtilde}
\eeq
where we have defined
\beq
\tilde\kappa = \frac{\kappa c_1}{\sqrt{c_0c_2}m}.
\eeq
The virial law is now given by
\beq
E'(\tR) = \frac{c_2^{\frac32}}{\sqrt{c_0}m}\left[1 - 2\tilde\kappa\tR + 3\tR^2\right] = 0.
\eeq
In contradistinction to the 2-dimensional magnetic Skyrmion case, in 3
dimensions there are two solutions to the virial equation:
\beq
R_\pm = \frac{\tilde\kappa\pm\sqrt{\tilde\kappa^2-3}}{3}R_0.
\label{eq:Rpm_magskyrm_sphaleron}
\eeq
Clearly, there are no solutions if $\tilde\kappa<\sqrt{3}$, in which
case the DM term is not strong enough to stabilize a soliton solution.
However, as soon as $\tilde\kappa$ is big enough, two solutions exist:
the large one is the 3D magnetic Skyrmion and we call the small one a
magnetic sphaleron.
Although we cannot expect $c_{2,1,0}$ to be truly constants for all
values of the parameters of the energy functional, we assume their
dependence to be mild.
Nevertheless, it is illuminating to expand in small $m$:
\begin{align}
R_- &= \frac{c_2}{2c_1\kappa} + \frac{3c_0c_2^2m^2}{8c_1^3\kappa^3} + \mathcal{O}(m^4), \label{eq:R_} \\ 
R_+ &= \frac{2c_1\kappa}{3c_0m^2} - \frac{c_2}{2c_1\kappa} - \frac{3c_0c_2^2m^2}{8c_1^3\kappa} + \mathcal{O}(m^4). \label{eq:R+}
\end{align}
The solution corresponding to $R_-$, which is the magnetic
sphaleron, tends to a finite size as $m$ tends to zero, whereas the
solution corresponding to $R_+$ diverges as $m\to0$.
The finite size of $R_-$ for $m=0$ shows that the massless magnetic
sphaleron is stabilized exclusively by the balance of the kinetic and
the DM term.
The latter solution, $R_+$, is the 3D magnetic Skyrmion.
Since $R_+$ diverges in the limit of $m\to0$, the Skyrmion disappears
in that limit.

\subsubsection{Numerical sphaleron solutions}

We solve the equation of motion \eqref{eq:eom_chi} with $\kappa$ set
to unity (under the rescaled scheme) using the shooting method.
The reason for changing the algorithm from gradient flow to the
shooting method,
is that the magnetic sphaleron solutions are unstable, which makes the
gradient flow method easily flow to the stable 3D magnetic Skyrmions
instead.
Linearizing the equation of motion \eqref{eq:eom_chi}, we find that at
small distances $r\ll 1$, the profile function behaves like
$\chi=\pi-c r+\mathcal{O}(r^2)$.
We use $c$ as the shooting parameter and require
$\chi(r_{\rm max})=0$, with a suitably large $r_{\rm max}$.
The typical values of the shooting parameter for the magnetic
sphalerons are in the range $c\approx[2,6]$.

\begin{figure}[!htp]
  \centering
  \mbox{\subfloat[$p=1$]{\includegraphics[width=0.49\linewidth]{{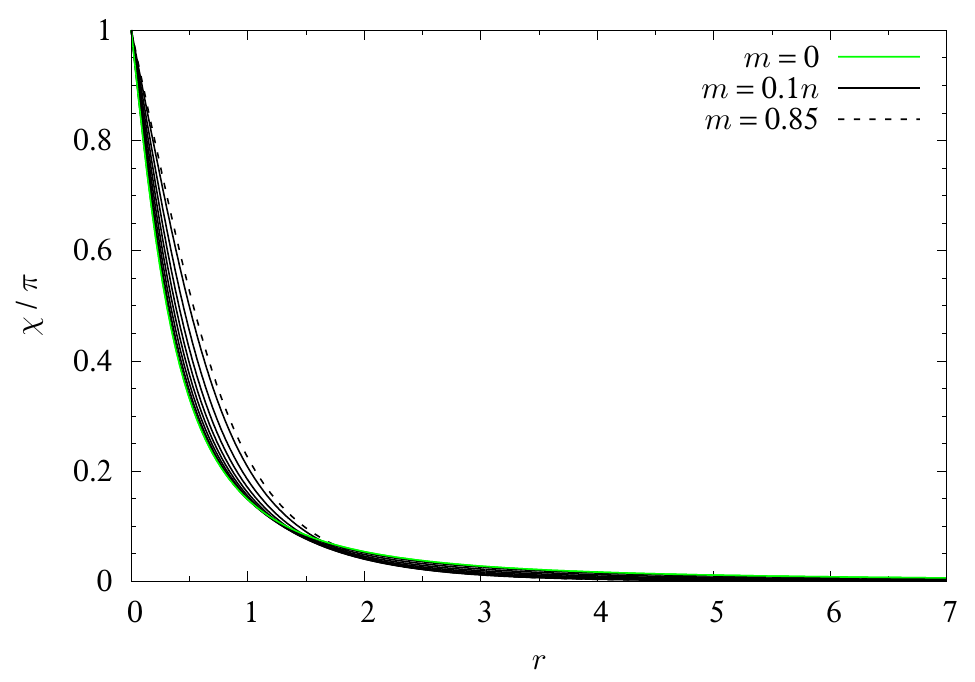}}}
      \subfloat[$p=\frac32$]{\includegraphics[width=0.49\linewidth]{{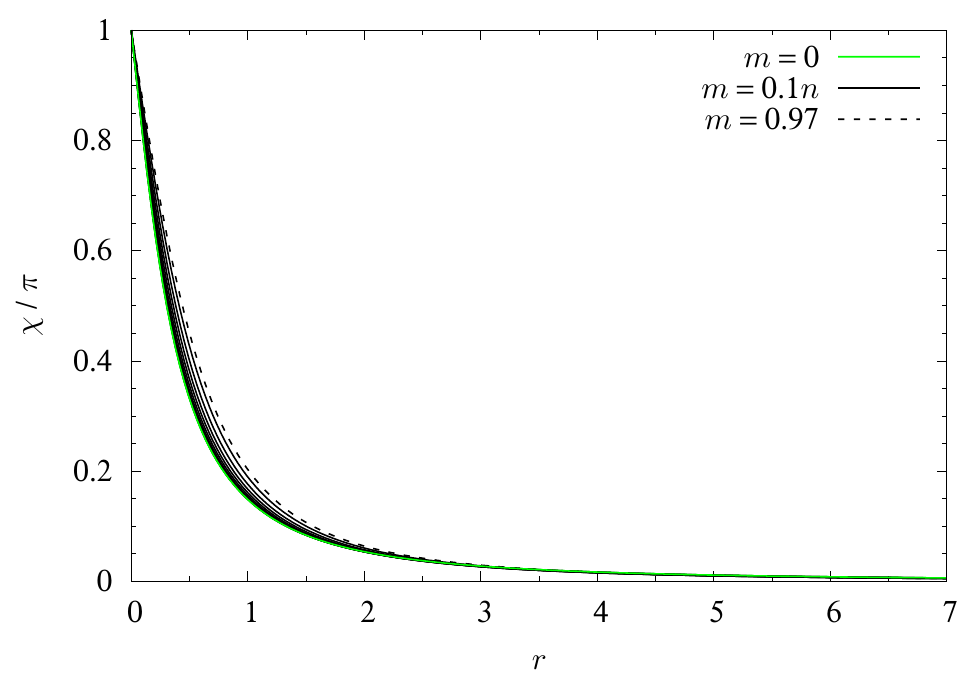}}}}
  \mbox{\subfloat[$p=2$]{\includegraphics[width=0.49\linewidth]{{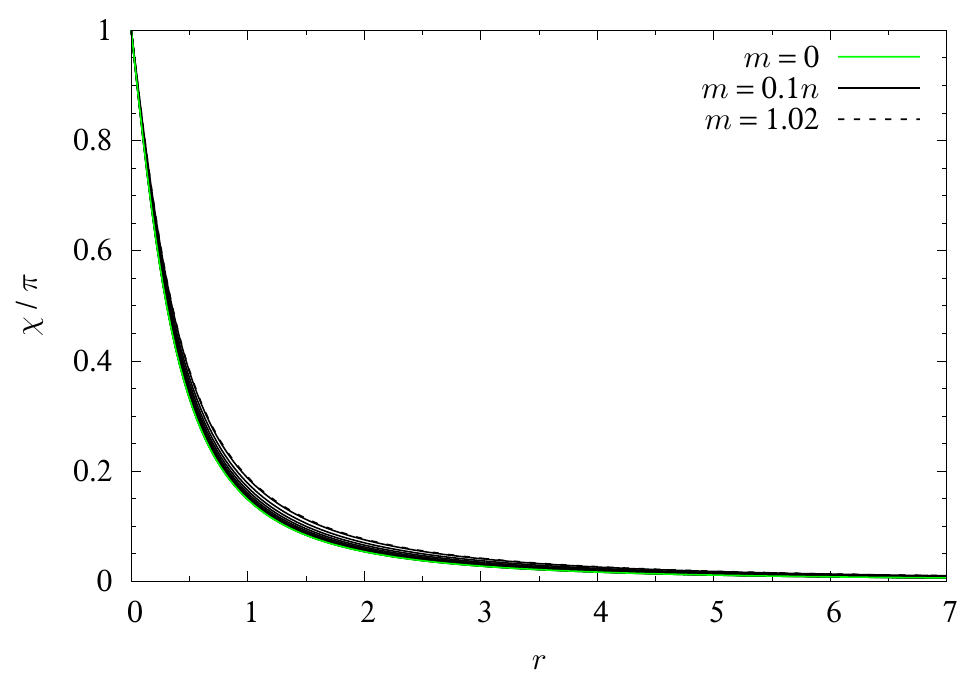}}}}
  \caption{Profile functions for magnetic sphalerons with higher-dimensional
    $\SO(3)_{\rm diag}$-invariant DM term, for (a) $p=1$, (b) $p=\tfrac32$ and
    (c) $p=2$.
  The $m=0$ solution is shown with a green line and the  $m_{\rm crit}$ solution with a dashed line.
  No solutions exist for $m>m_{\rm crit}$.
  }
  \label{fig:sphaleron_sols}
\end{figure}
Fig.~\ref{fig:sphaleron_sols} shows the profile functions of the
solutions for the full range of parameters,
i.e.~$m\in[0,m_{\rm crit}]$ in the rescaled lengths where $\kappa$
only enters as the (inverse) unit of energy.
As in the 3D magnetic Skyrmion case, there exist no
solution for $m>m_{\rm crit}$, as can also be seen from
Eq.~\eqref{eq:Rpm_magskyrm_sphaleron}.
The upper bound on $m$ corresponds to the lower bound on
$\tilde\kappa\geq\sqrt{3}$ and when satisfied, both the 3D magnetic
Skyrmion and the magnetic sphaleron exist.

Interestingly, we can see that increasing the potential (mass)
parameter, $m$, inflates the bulk of the soliton but suppresses the
tail for $p=1$. For $p>1$, the tail is not suppressed in the same way,
since the potential is not simply giving rise to an exponential
behavior, but is nonlinearly suppressing the tail.

\begin{figure}[!htp]
  \centering
  \mbox{\subfloat[$p=1$]{\includegraphics[width=0.4\linewidth]{{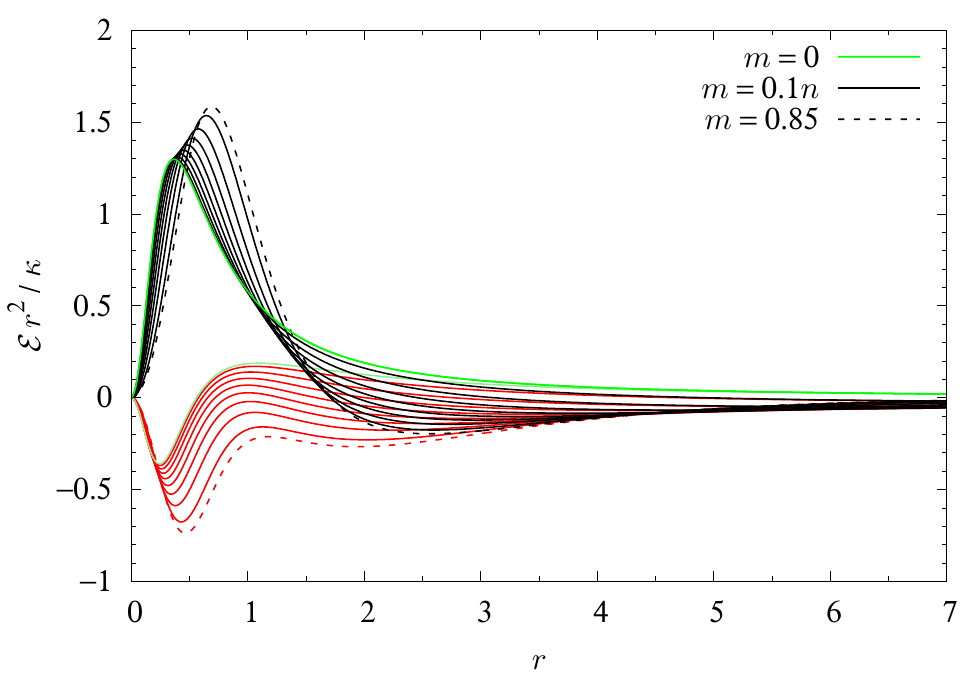}}}
    \subfloat[$p=1$]{\includegraphics[width=0.4\linewidth]{{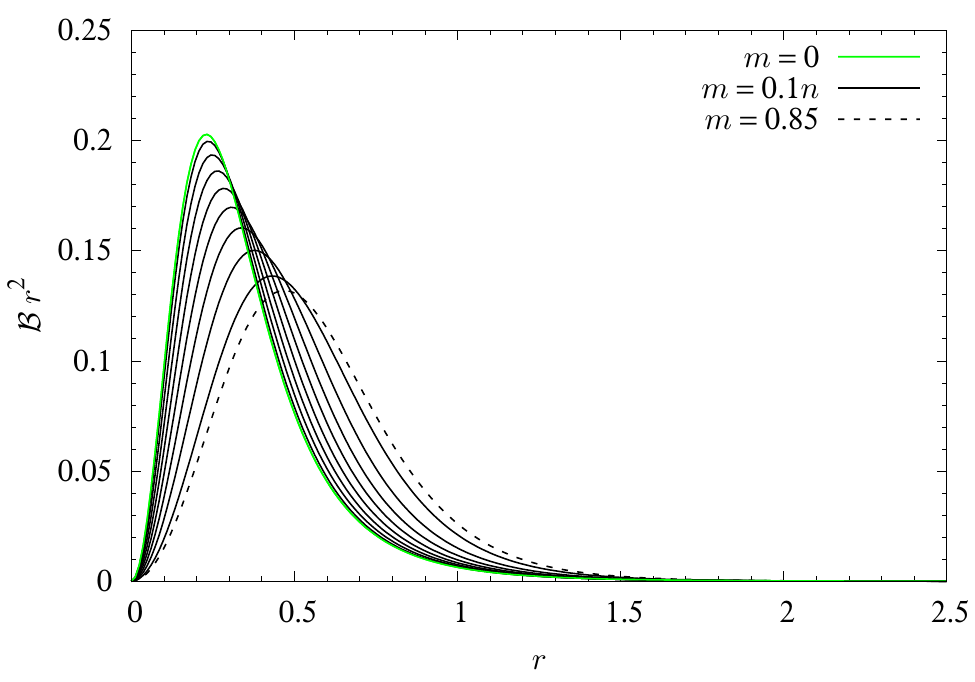}}}}
  \mbox{\subfloat[$p=\frac32$]{\includegraphics[width=0.4\linewidth]{{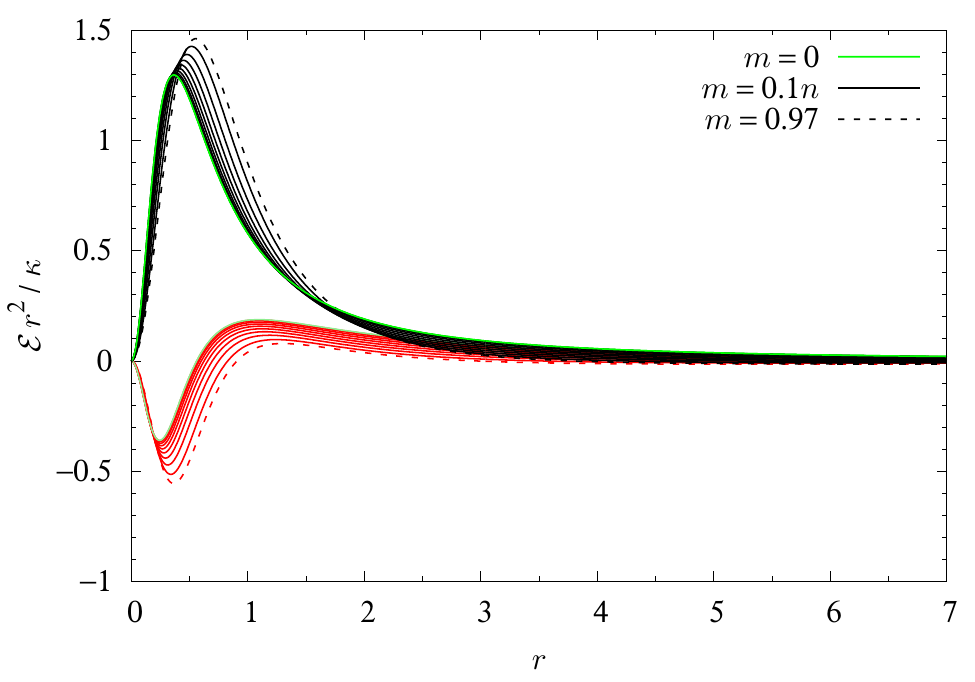}}}
    \subfloat[$p=\frac32$]{\includegraphics[width=0.4\linewidth]{{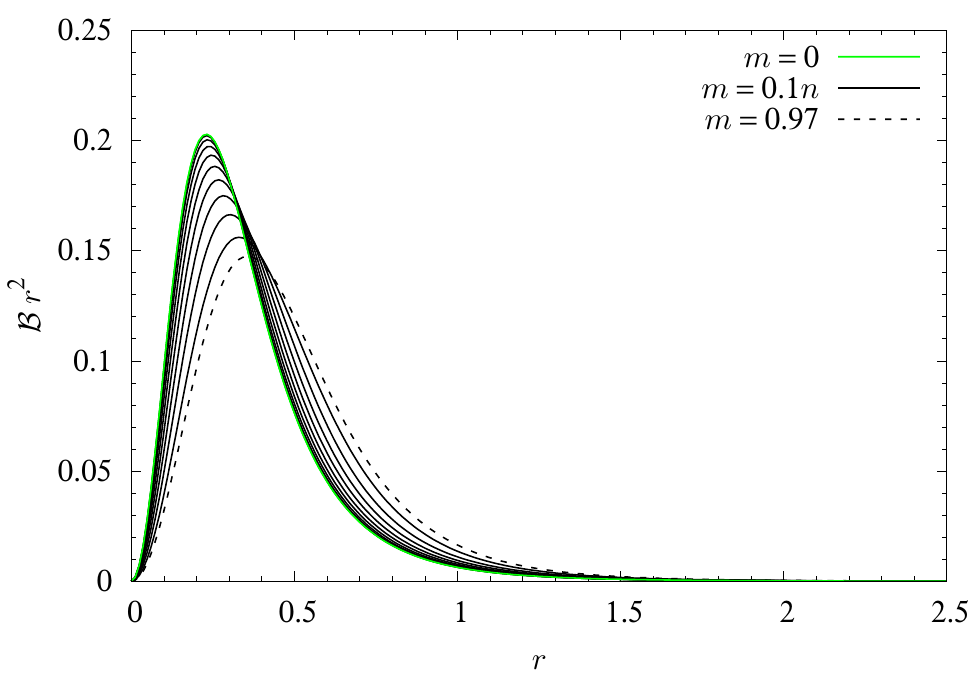}}}}
  \mbox{\subfloat[$p=2$]{\includegraphics[width=0.4\linewidth]{{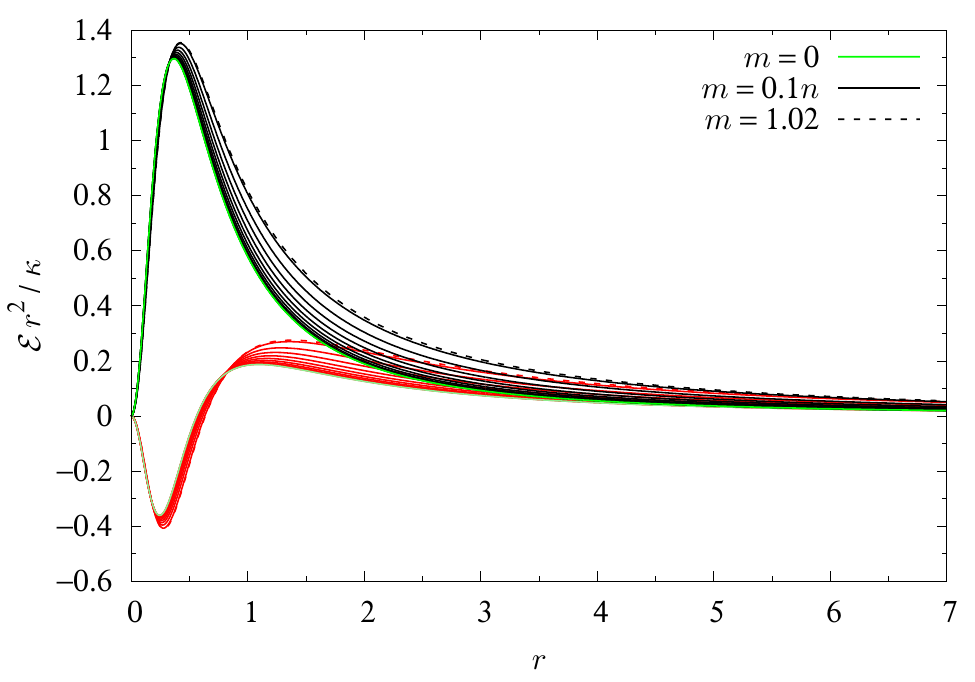}}}
    \subfloat[$p=2$]{\includegraphics[width=0.4\linewidth]{{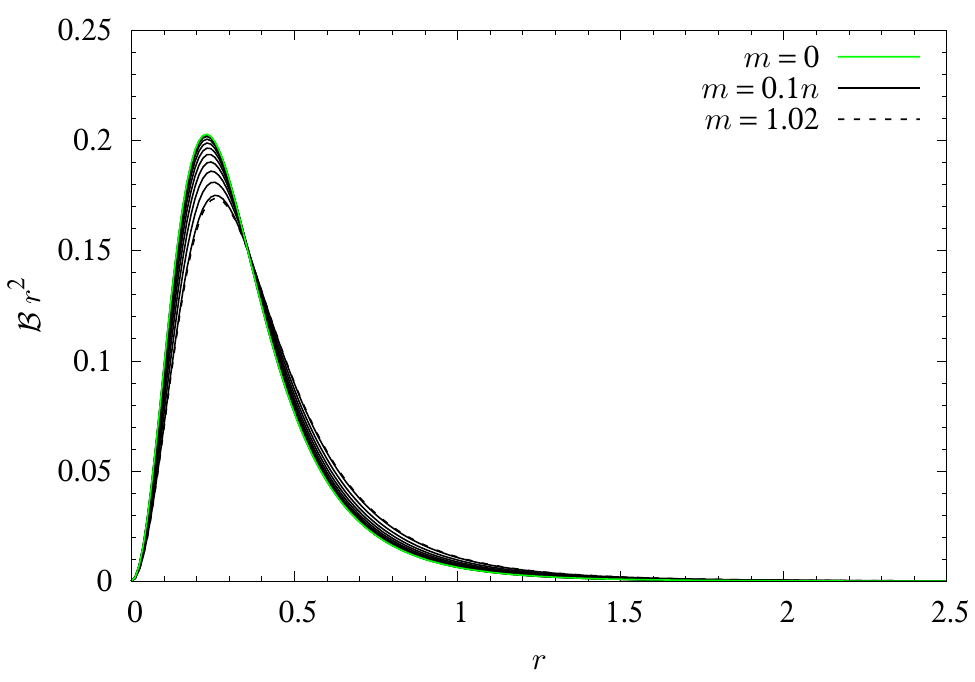}}}}
  \caption{Energy density and topological charge density for magnetic
    sphalerons with higher-dimensional $\SO(3)_{\rm diag}$-invariant DM term.
  Left panels (a,c,e) show the total energy densities weighted with
  $r^2$ as black lines and
  the DM contribution as red lines. Right panels
  (b,d,f) show the topological charge density, weighted with $r^2$.
  The $m=0$ solution is shown with a green line (light-green for the DM
  term) and the $m_{\rm crit}$ solution is shown with
  a dashed line.
  The rows of panels correspond to the potential power parameter
  $p=1,\tfrac32,2$.
  }
  \label{fig:sphaleron_ensbs}
\end{figure}
Fig.~\ref{fig:sphaleron_ensbs} shows the energy densities and
topological charge densities.
The left panels show the total energy densities with black lines and
the DM contribution with red lines.
The 3 rows of the figure correspond to different values of the
potential power parameter, $p=1,\tfrac32,2$.

\begin{figure}[!htp]
  \centering
  \mbox{\subfloat[]{\includegraphics[width=0.49\linewidth]{{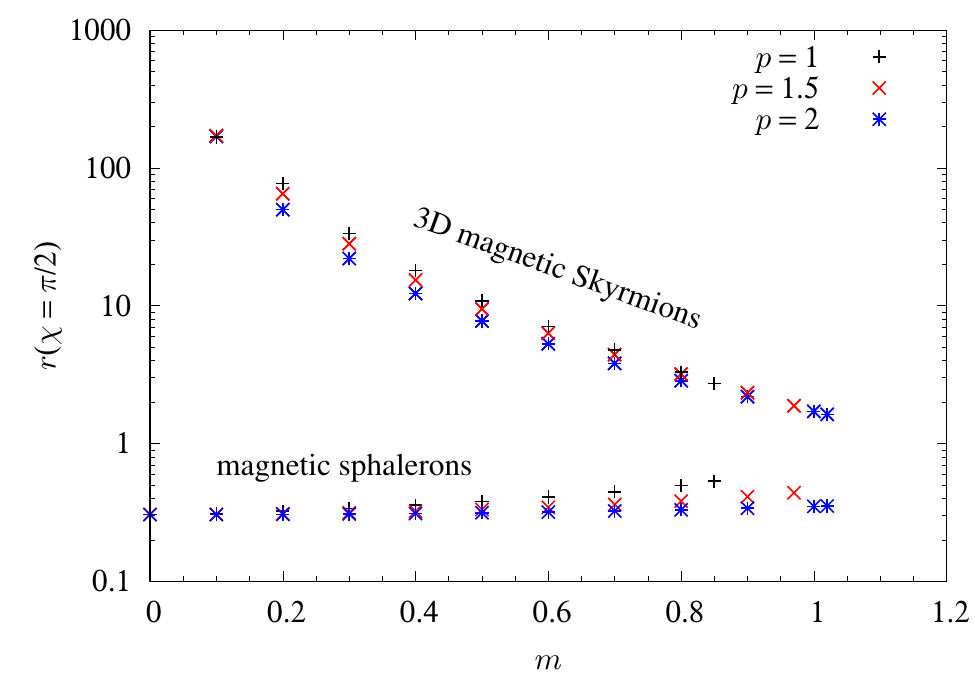}}}
  \subfloat[]{\includegraphics[width=0.49\linewidth]{{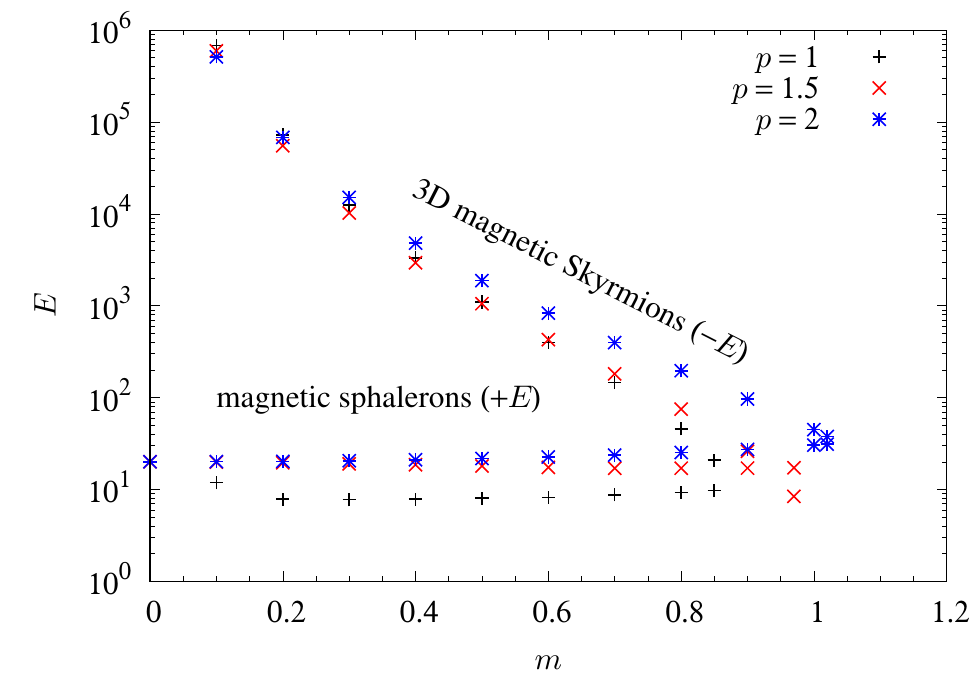}}}}
  \caption{(a) Sizes and (b) energies of magnetic sphalerons (bottom
    series) versus 3D magnetic Skyrmions (top series) with higher-dimensional
    $\SO(3)_{\rm diag}$-invariant DM term.
    (a) The size is measured simply by the radius where
    $\chi=\frac\pi2$.
    (b) $-E$ is shown for the magnetic Skyrmions, as they have
    negative energy (i.e.~they are stable solitons).
  Black pluses correspond to $p=1$, red crosses to $p=\tfrac32$ and
  blue plus-crosses to $p=2$.
  The critical masses, $m_{\rm crit}$, correspond to the largest 
  sphalerons and the smallest Skyrmions.
  }
  \label{fig:size}
\end{figure}

The sizes of the Skyrmions given in Fig.~\ref{fig:size}(a) are found
simply by determining the radius where the profile function $\chi$ is
half-way between the anti-vacuum and the vacuum,
i.e.~$\chi(r)=\tfrac\pi2$.
The energies are shown in Fig.~\ref{fig:size}(b) with positive sign
for the sphaleron and $-E$ is shown for the magnetic Skyrmion (since
it is a logarithmic plot).
Clearly, the sphaleron has a larger energy than the magnetic Skyrmion,
since it always has the opposite sign than the latter.

\begin{figure}[!htp]
  \centering
  \mbox{\subfloat[]{\includegraphics[width=0.49\linewidth]{{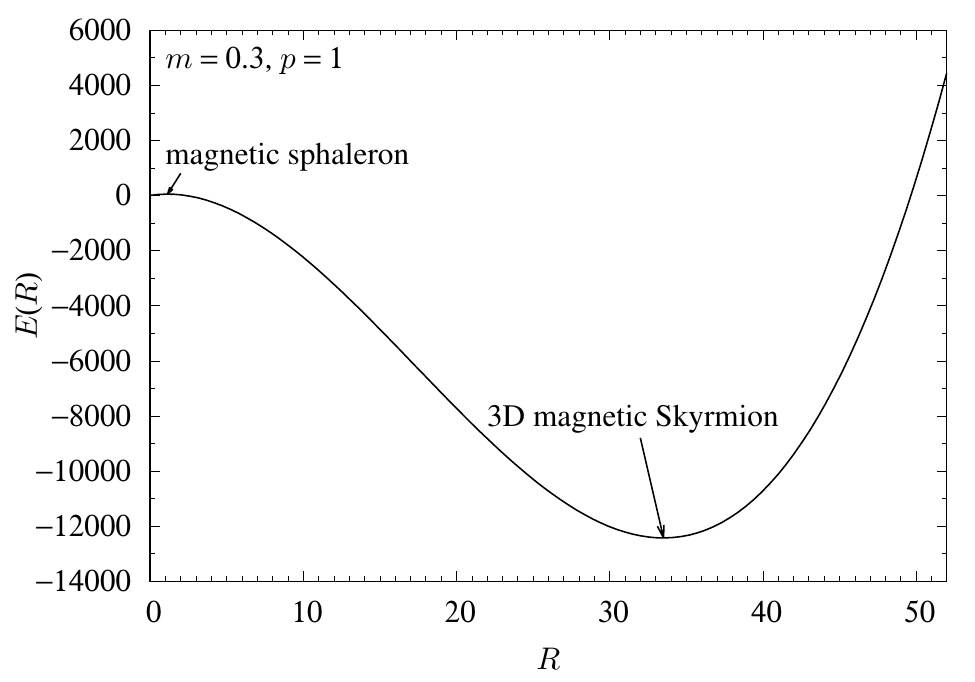}}}
  \subfloat[]{\includegraphics[width=0.49\linewidth]{{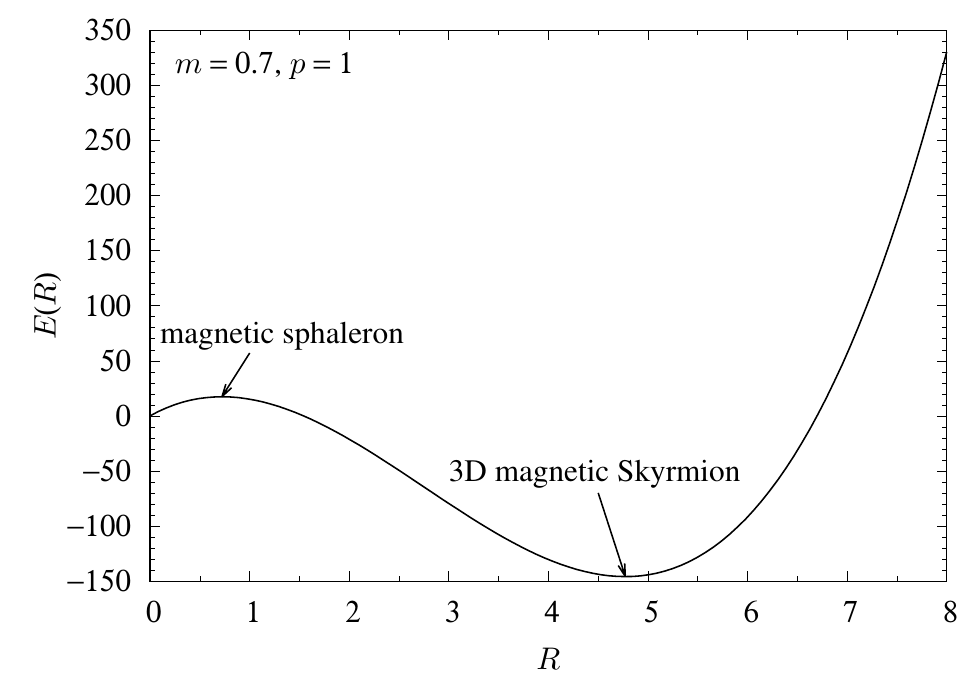}}}}
  \caption{The energy functional $E(R)$ as function of the radius $R$,
    illustrating the unstable magnetic sphaleron fixed point at a small radius
    and the stable magnetic Skyrmion fixed point at a larger radius,
    for (a) $m=0.3$ and (b) $m=0.7$ both with the $p=1$ potential.
    It is also observed that the magnetic sphaleron grows with $m$
    whereas the magnetic Skyrmion shrinks with $m$.
  }
  \label{fig:ER}
\end{figure}
Finally, in Fig.~\ref{fig:ER} we show the energy functional $E(R)$ of
Eq.~\eqref{eq:ER} for the two cases of $m=0.3$ and $m=0.7$ with the
potential $p=1$.
It is observed from the figure that the magnetic sphaleron has a positive
energy, whereas the magnetic Skyrmion always has a negative energy.
Moreover, the scaling of the radii is consistent with
Fig.~\ref{fig:size}; that is, the magnetic sphaleron grows while the
magnetic Skyrmion shrinks with increasing $m$, consistent with Eqs.~\eqref{eq:R_} and~\eqref{eq:R+} .

\subsection{A connection to the Hopfion}\label{sec:magSkHopf}

At present, the 3-dimensional chiral magnetic Skyrmion we proposed
stands as a purely theoretical model with no experimental
realization, since we require a 4-dimensional magnetization vector.
However, our higher-dimensional model can be connected to the
model of Hopfions in chiral magnets which have been studied theoretically in
Refs.~\cite{tai2018static,liu2018binding,sutcliffe2018hopfions} and
found experimentally \cite{ackerman2017static,kent2021creation}.
Magnetic Hopfions have also been studied in frustrated
magnets in Ref.~\cite{sutcliffe2017skyrmion,sutcliffe2018hopfionsReview}.
In order to interpolate to the Hopfion model, we need to turn on an
additional deformation potential
\beq
E_0^{\rm def}[\bm{n}]=\int\d^3x\;{m'}^2(n^3)^2,
\label{eq:deformation}
\eeq
so that the theory is $E[\bm{n}]+E_{0}^{\rm def}[\bm{n}]$, with
$E[\bm{n}]$ of Eq.~\eqref{Emagsky}.
This potential is compatible with the vacuum $\bm{N}=(0,0,0,1)^{\rm T}$, but
squeezes the unbroken 2-sphere.
In the limit $m'\to0$ we return to our higher-dimensional model,
whereas in the limit $m'\to\infty$, the target space is effectively
reduced from the 3-dimensional $S^3$ to the 2-dimensional $S^2$.
Being too energetically expensive to deviate from $n^3=0$, the
theory with the $\SO(3)_{\rm diag}$-invariant higher-dimensional DM term
\eqref{eq:Theta_inv_standard} reduces in the $m'\to\infty$ limit to
\beq
E^{\rm red}[\bm{u}]=\int\d^3x\;\left[
  \frac12\p_i\bm{u}\cdot\p_i\bm{u}
  +\kappa\alpha\bm{u}\cdot\nabla\times\bm{u}
  -\kappa(\beta-\alpha)\epsilon^{ij}u^i\p_3u^j
  +m^2(1-u^3)
  \right],
\eeq
where $\bm{u}=(-n^2,n^1,n^4)\in S^2$ is the reduced magnetization
vector ($\bm{u}\cdot\bm{u}=1$).
Notice the rotation of the first two magnetization coordinates.
Note also that the $\p_3$ derivative part of the DM term is
proportional to $\beta$ only.
Setting $\beta=\alpha=1$ simplifies the energy to
\beq
E^{\rm red}[\bm{u}]=\int\d^3x\;\left[
  \frac12\p_i\bm{u}\cdot\p_i\bm{u}
  +\kappa\bm{u}\cdot\nabla\times\bm{u}
  +m^2(1-u^3)
  \right],
\eeq
which is exactly the energy of the Hopfion studied in
Refs.~\cite{tai2018static,liu2018binding,sutcliffe2018hopfions}.

Given that the Derrick's scaling of this theory is identical to our
higher-dimensional model with $\bm{n}\in S^3$ instead of $\bm{u}\in
S^2$, we expect that also the magnetic Hopfion model in chiral magnets
could have a \emph{sphaleron}, see Sec.~\ref{sec:sphalerons}.

A problem with the deformation \eqref{eq:deformation} is that the
hybrid Skyrmion sector (with $S^3$ target space) and the Hopfion
sector (with $S^2$ target space) are not connected continuously.
Sending $m'\to\infty$ the Skyrmion mass in the nontrivial sector
$\pi_3(S^3)$ goes to infinity.
The Hopfion sector, $\pi_3(S^2)$, instead appears at a certain $m'$ as
a metastable sector and becomes absolutely stable as $m'\to\infty$.

\section{A magnetic-QCD hybrid Skyrme model}\label{sec:hybrid}

We will now briefly consider the hybrid model consisting of the
higher-dimensional magnetic Skyrmion model of Sec.~\ref{sec:3D} and
the standard Skyrme model that is often used in a high-energy context
\cite{skyrme1961non,skyrme1962unified}.
We do this by amending the energy~\eqref{Emagsky} with the so-called Skyrme
term, $E_4 = \int \d^3x\,\E_4$, whose density is \cite{skyrme1961non,skyrme1962unified}
\beq
\E_4(\bm{n}) = \frac{1}{4e^2}\big[
  (\p_i\bm{n}\cdot\p_i\bm{n})(\p_j\bm{n}\cdot\p_j\bm{n})
  -(\p_i\bm{n}\cdot\p_j\bm{n})(\p_i\bm{n}\cdot\p_j\bm{n})\big].
  \label{skyrme_density}
\eeq
This introduces an extra coupling constant or theory parameter,
viz.~$e$ -- not to be confused with the electron charge.

Inserting the hedgehog Ansatz \eqref{eq:hedgehog}, the total energy
for the magnetic-QCD hybrid Skyrme model reads
\begin{align}
  E &= 4\pi\int\bigg[
      \frac12r^2(\chi')^2
  +\sin^2\chi
  +\kappa(r^2\chi' + r\sin2\chi)
  +m^2 r^2(1-\cos\chi)^p\non
  &\phantom{=4\pi\int\Bigg[\ }
    +\frac{\sin^2\chi}{e^2}\left((\chi')^2 + \frac{\sin^2\chi}{2r^2}\right)
  \bigg] \d r,
\end{align}
which leads to the equation of motion
\beq
\chi''
+ \frac{2}{r}\chi'
- \frac{\sin2\chi}{r^2}
+ \frac{4\kappa\sin^2\chi}{r}
- p m^2(1-\cos\chi)^{p-1}\sin\chi\qquad\non
\mathop+ \frac{2\sin^2(\chi)\chi''}{e^2r^2}
+ \frac{\sin(2\chi)(\chi')^2}{e^2r^2}
- \frac{\sin^2\chi\sin2\chi}{e^2r^4}
= 0.
\label{eq:eom_hybrid}
\eeq
The DM term must be negative to stabilize the soliton according to
Derrick's theorem.
If we add a boundary term $2\pi\kappa\p_r(r^2\sin2\chi)$ to
the energy functional, the DM term becomes negative definite for a
monotonically decreasing profile function:
\begin{align}
  E &= 4\pi\int\bigg[
      \frac12r^2(\chi')^2
  +\sin^2\chi
  +\kappa r^2\sin^2(\chi)\chi'
  +m^2 r^2(1-\cos\chi)^p\non
  &\phantom{=4\pi\int\Bigg[\ }
    +\frac{\sin^2\chi}{e^2}\left((\chi')^2 + \frac{\sin^2\chi}{2r^2}\right)
  \bigg] \d r,
\end{align}
that is, for $\chi(r)$ going monotonically from $\chi(0)=\pi$ to
$\chi(\infty)=0$.
We should note that the boundary term only vanishes for exponentially
localized solitons ($m\neq 0$), otherwise for massless Skyrmions this
boundary term will give a finite, but constant contribution to the energy. 

We are now ready to perform a Derrick's scaling analysis next.

\subsection{Derrick phase diagram}\label{sec:hybrid_Derrick_scaling}

In this section, we investigate the phase diagram of the theory
through Derrick's theorem.
Rescaling the lengths as $r\to\lambda r$, we obtain similarly to the
case of the model without the Skyrme term, a scaled energy
\beq
E_\lambda = \lambda E_2 + \lambda^2 E_1 + \lambda^3 E_0 + \frac{1}{\lambda}E_4.
\eeq
Each energy contribution is here approximated to be a constant $c_n$
times the coupling constant times the soliton size $R$ to the power
given by the Derrick scaling:
\beq
\lambda E_2 = c_2 R, \qquad
\lambda^2 E_1 = -c_1\kappa R^2, \qquad
\lambda^3 E_0 = c_0 m^2 R^3, \qquad
\frac{1}{\lambda}E_4 = \frac{c_4}{e^2R}.
\eeq
This approximate energy functional as a function of the soliton size,
$R$, thus reads
\begin{equation}
  E(R) =  c_2 R - \kappa c_1 R^2 + m^2 c_0 R^3 + \frac{c_4}{e^2R} \ .
  \label{erpol2}
\end{equation}
At this point, it will prove useful to choose a particular unit of
energy and of length, so as to reduce the variables of the theory.
In particular, we have
\beq
E(\tR) = \frac{c_2^{\frac32}}{\sqrt{c_0}m}\left[
  \tR - \tilde{\kappa}\tR^2 + \tR^3 + \frac{1}{\tilde{e}^2\tR}
  \right],
\eeq
where we have defined
\beq
\tilde{\kappa}=\frac{\kappa c_1}{\sqrt{c_0c_2}m},\qquad
\tilde{e}=\frac{e c_2}{\sqrt{c_0c_4}m},
\eeq
and $\tR$ is given in Eq.~\eqref{eq:Rtilde}

Derrick stability is given by imposing the derivative of the above energy with respect
to $R$ equal to zero:
\beq
E'(\tR) = \frac{c_2^{\frac32}}{\sqrt{c_0}m}\left[
  1 - 2\tilde\kappa \tR + 3\tR^2 - \frac{1}{\tilde{e}^2\tR^2}
  \right]
  = 0 \ .
\eeq
This is a polynomial equation in $\tR$ of fourth order,
\beq
y(\tR) = \tR^4
-\frac{2\tilde\kappa}{3} \tR^3
+\frac{1}{3} \tR^2
-\frac{1}{3\tilde{e}^2} = 0 \ ,
\eeq
which thus has in principle four roots.
Physically, only real and positive roots are relevant though.
Deriving this polynomial function with respect to $\tR$ gives the saddle points:
\begin{equation}
4\tR^2 - 2\tilde\kappa \tR + \frac{2}{3} = 0,
\quad\Rightarrow\quad
\tR^{\rm saddle}_\pm = \frac14\left(\tilde\kappa \pm \sqrt{\tilde\kappa^2-\frac83}\right),
\end{equation}
as well as $\tR=0$.
There are now four possibilities, if the argument of the square root is
positive, there are two saddle points, $\tR^{\rm saddle}_\pm$ in
addition to that at $\tR=0$.
We assume that $1/\tilde{e}>0$, i.e.~that the Skyrme term is turned on
(if not, the analysis of Sec.~\ref{sec:sphalerons} applies).
If $y(\tR_-^{\rm saddle})$ and $y(\tR_+^{\rm saddle})$ have the same
sign, then there is only a single solution.
If $y(\tR_-^{\rm saddle})>0$ and $y(\tR_+^{\rm saddle})<0$, there will
be three different solutions.
The third possibility is that $y(\tR_-^{\rm saddle})=0$ in which case
there will be exactly two solutions.
The fourth possibility that $y(\tR_+^{\rm saddle})=0$ also yields two
solutions.
Finally, for $\tilde\kappa<\sqrt{8/3}$ the saddle points are not real
and there is again only a single solution.
There is also only one solution for $\kappa=\sqrt{8/3}$, since
$\tR_-^{\rm saddle}$ coalesces with $\tR=0$ and the polynomial will
only cross $y=0$ once, even if that happens at the saddle point
$\tR_+^{\rm saddle}$.

Evaluating the polynomial $y$ at the saddle points gives
\begin{align}
y(\tR_\pm^{\rm saddle}) =
-\frac{1}{36}
-\frac{1}{3\tilde{e}^2}
+\frac{1}{864}\left(
36\tilde\kappa^2
-9\tilde\kappa^4
\pm8\tilde\kappa\sqrt{9\tilde\kappa^2-24}
\mp3\tilde\kappa^3\sqrt{9\tilde\kappa^2-24}
\right).
\end{align}
We notice, that if $\frac{1}{\tilde{e}^2}$ is too large, both saddle
points are negative and only one solution exists.
Setting $\kappa=\sqrt{8/3}$, the two saddle points coalesce.
In this case, the condition $y(\tR_-^{\rm saddle})>0$ becomes
\beq
\frac{1}{\tilde{e}^2} \leq \frac{1}{36},
\eeq
with equality at the triple point of the phase diagram.
On the other hand, the largest power of $\tilde\kappa$ in
$y(\tR_+^{\rm saddle})$ has a negative coefficient, so
$y(\tR_+^{\rm saddle})<0$ is always satisfied for large
$\tilde\kappa$.
The condition $y(\tR_-^{\rm saddle})\geq0$ for large $\tilde\kappa$ can
be simplified by Taylor expansion to
\beq
\frac{1}{\tilde{e}^2} \leq \frac{1}{27\tilde\kappa^2} + \mathcal{O}(\tilde\kappa^{-4}).
\eeq
\begin{figure}[!htp]
  \centering
  \includegraphics[width=0.5\linewidth]{{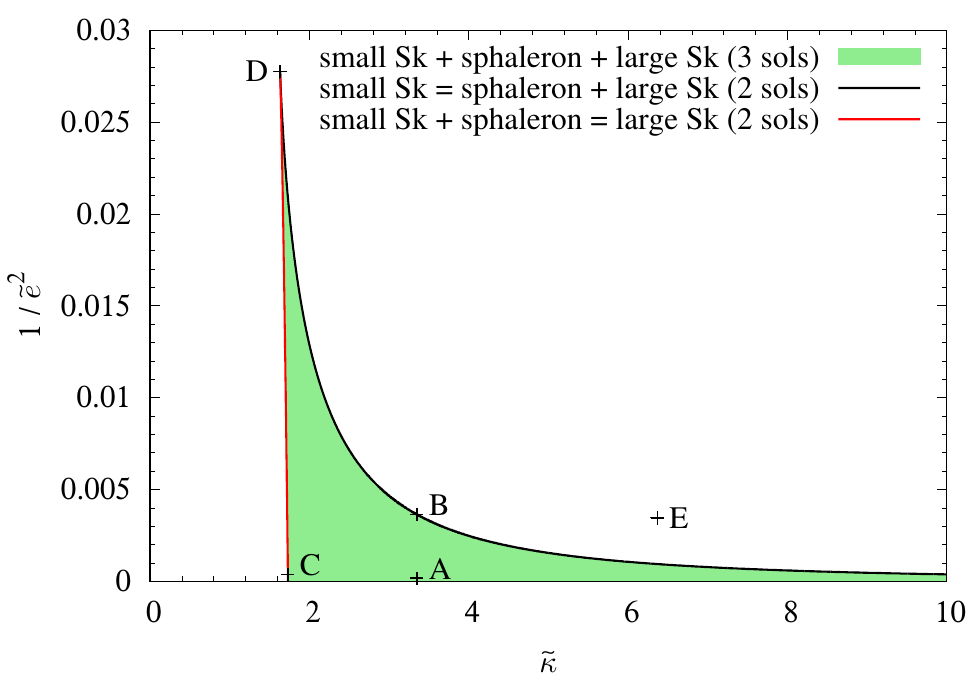}}
  \caption{The green shaded area of the phase diagram contains 3
    solutions (a small Skyrmion, a sphaleron and a large Skyrmion),
    the black line corresponds to $y(\tR_-^{\rm saddle})=0$ for which
    the small Skyrmion and the sphaleron coalesce, and the red line corresponds to
    $y(\tR_+^{\rm saddle})=0$ for which the sphaleron and the large
    Skyrmion coalesce.
    In addition there is only 1 solution in the entire white area of
    the phase diagram.
    `Sk' in the legend is an abbreviation for Skyrmion.
  }
  \label{fig:Derrick_phase}
\end{figure}
The possibility for 3 solutions, corresponding to a small Skyrmion, a
big Skyrmion and a sphaleron, only exists in a narrow green shaded
band shown in Fig.~\ref{fig:Derrick_phase}.
The black line in the figure corresponds to $y(\tR_-^{\rm saddle})=0$,
for which there are two solutions: the small Skyrmion and the
sphaleron merge.
The red line corresponds to $y(\tR_+^{\rm saddle})=0$, for which there
are also two solutions: the sphaleron and the large Skyrmion merge.
It is difficult to see with the naked eye, but the red line is not
vertical, but slightly tilted, i.e.~it has a large negative derivative
with respect to $\tilde\kappa$.
The triple point of the phase diagram where the black line and the red
line meet is at
$(\tilde\kappa,\tilde{e}^{-2})=(\sqrt{8/3},1/36)$, for which there is
only one solution.
In the remaining white area of the phase diagram there is only a
single solution.

\subsection{Numerical magnetic-QCD hybrid Skyrmion solutions}

We will now turn to numerical solutions of the equation of motion
\eqref{eq:eom_hybrid} for the magnetic-QCD hybrid Skyrme model.
Considering mainly the aspect of the different points in the phase
diagram having a different number of solutions, we select 5
characteristic points in the phase diagram labeled as A through E, see
Fig.~\ref{fig:Derrick_phase}.

\begin{figure}[!htp]
  \centering
  \mbox{\subfloat[]{\includegraphics[width=0.49\linewidth]{{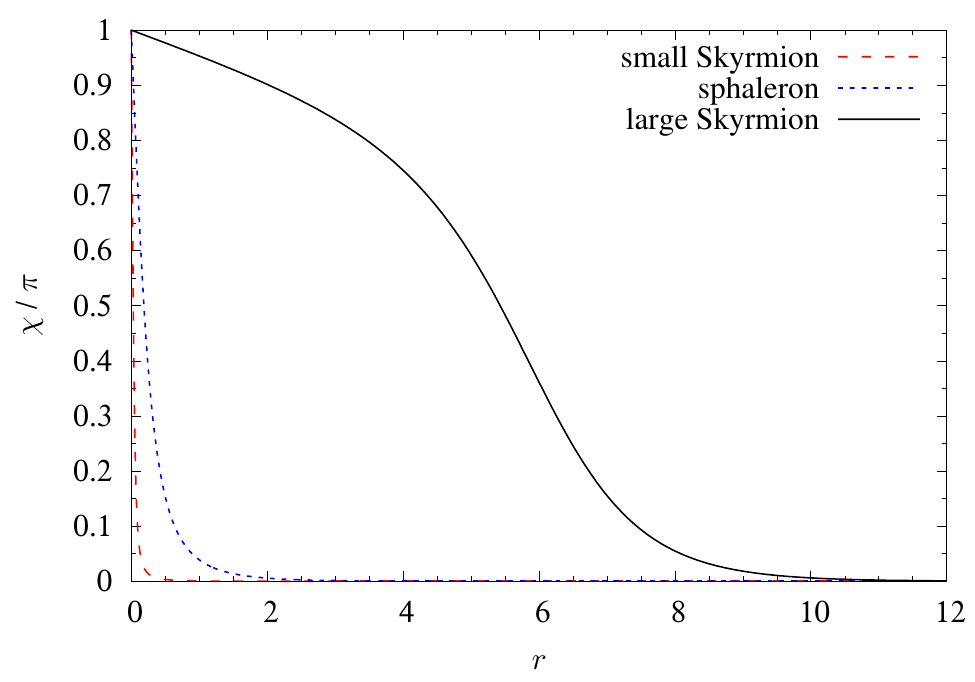}}}
    \subfloat[]{\includegraphics[width=0.49\linewidth]{{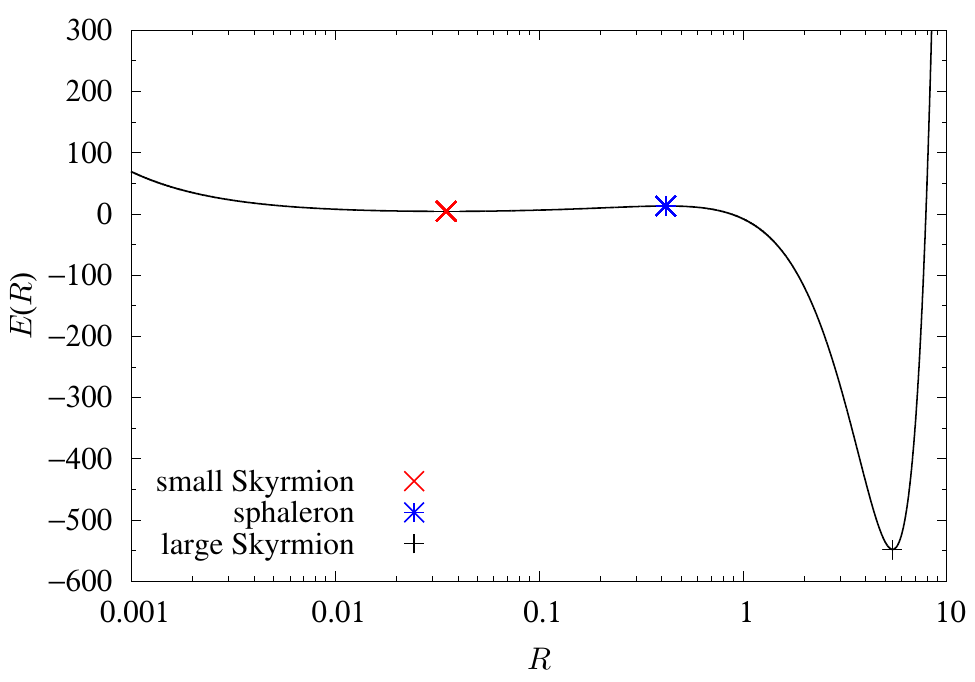}}}}
  \caption{(a) Three different solutions in the magnetic-QCD hybrid Skyrme
  model at the point in parameter space corresponding to A in the
  phase diagram \ref{fig:Derrick_phase}: the small Skyrmion, the
  sphaleron and the large Skyrmion; which are metastable, unstable and
  stable, respectively, as can be seen in panel (b).
  (b) The energy as function of $R$, see Eq.~\eqref{erpol2}.
  In this figure $\kappa=2$ and $1/e^2=0.001$ corresponding to
  $\tilde\kappa=3.34854$ and $1/\tilde{e}^2=0.000163315$. }
  \label{fig:hybrida}
\end{figure}
\begin{figure}[!htp]
  \centering
  \mbox{\subfloat[]{\includegraphics[width=0.49\linewidth]{{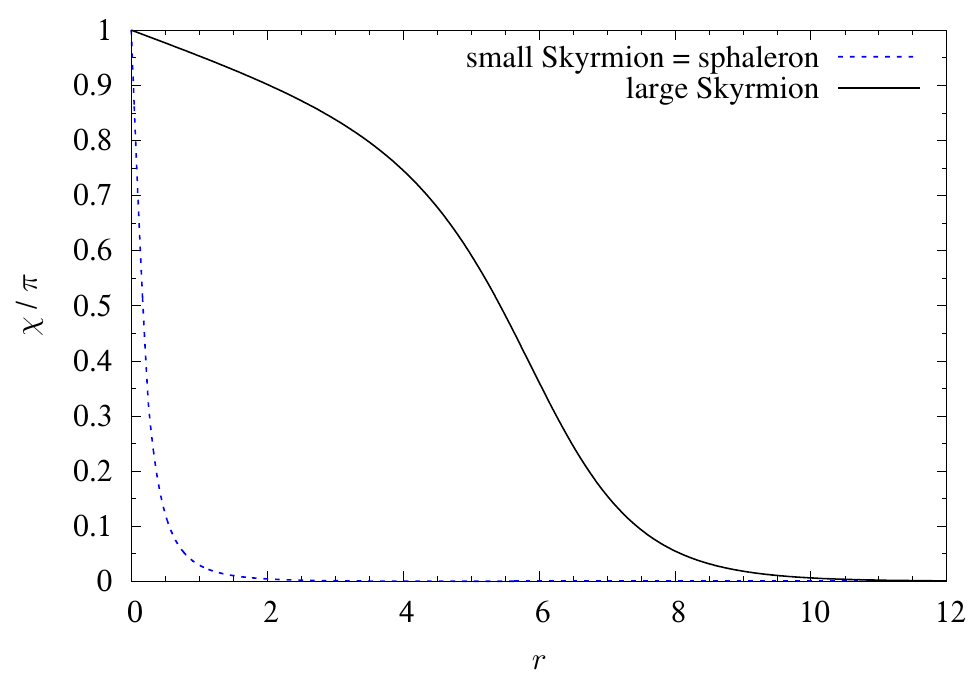}}}
    \subfloat[]{\includegraphics[width=0.49\linewidth]{{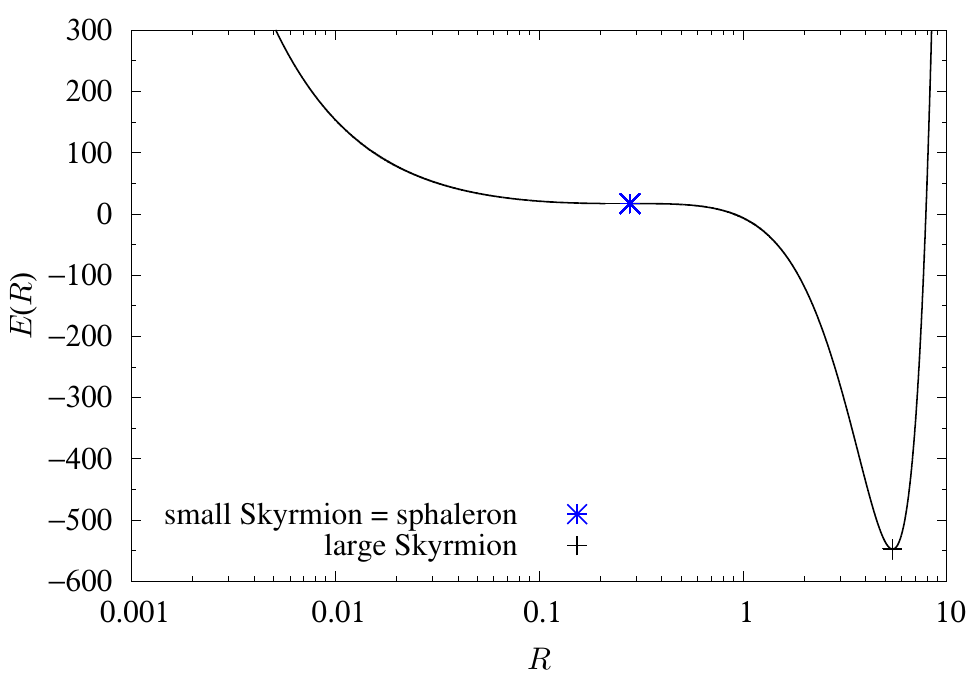}}}}
  \caption{(a) Two different solutions in the magnetic-QCD hybrid Skyrme
  model at the point in parameter space corresponding to B in the
  phase diagram \ref{fig:Derrick_phase}: the
  sphaleron and the large Skyrmion; which are unstable and
  stable, respectively, as can be seen in panel (b).
  The small Skyrmion and the sphaleron have merged, as this point of
  the parameter space is on the black line of
  Fig.~\ref{fig:Derrick_phase}. 
  (b) The energy as function of $R$, see Eq.~\eqref{erpol2}.
  In this figure $\kappa=2$ and $1/e^2=0.0222714$ corresponding to
  $\tilde\kappa=3.34885$ and $1/\tilde{e}^2=0.00363763$. }
  \label{fig:hybridb}
\end{figure}
\begin{figure}[!htp]
  \centering
  \mbox{\subfloat[]{\includegraphics[width=0.49\linewidth]{{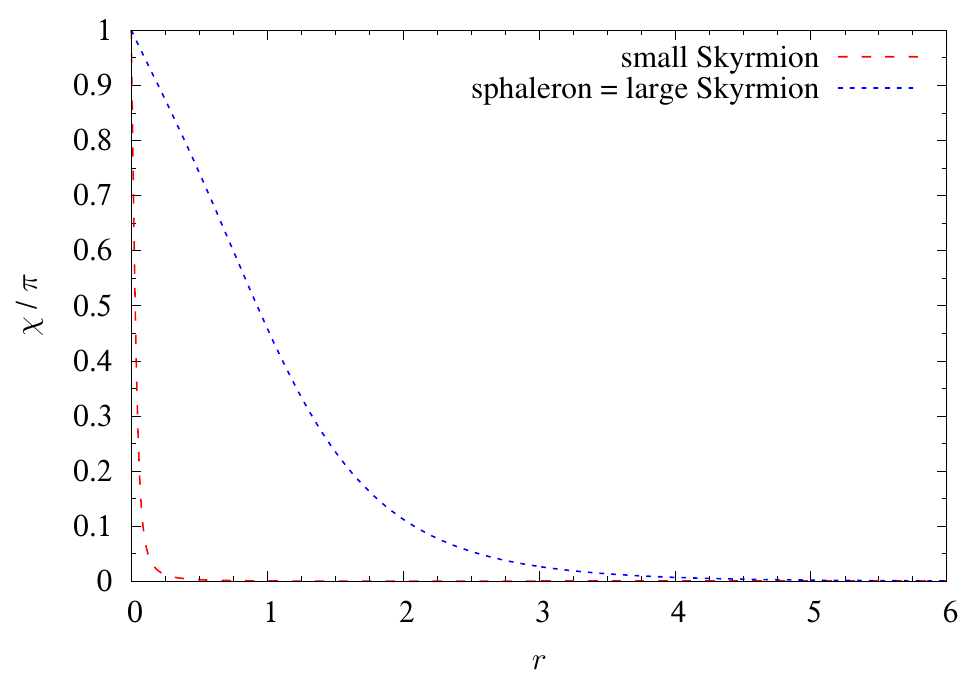}}}
    \subfloat[]{\includegraphics[width=0.49\linewidth]{{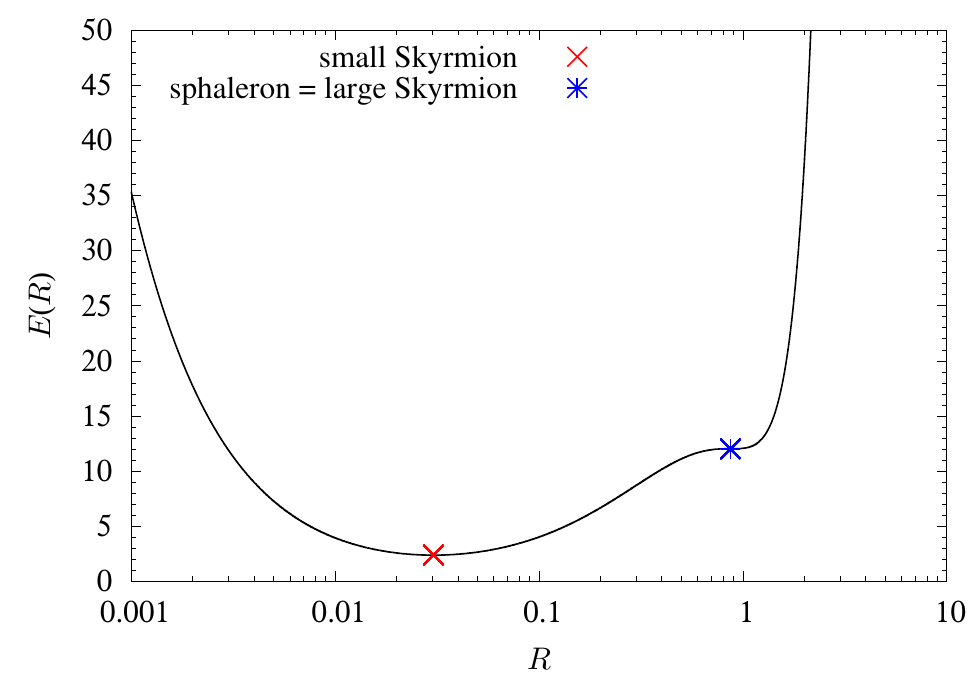}}}}
  \caption{(a) Two different solutions in the magnetic-QCD hybrid Skyrme
  model at the point in parameter space corresponding to C in the
  phase diagram \ref{fig:Derrick_phase}: the small Skyrmion and the
  sphaleron; which are stable and
  unstable, respectively, as can be seen in panel (b).
  The sphaleron and the large Skyrmion have merged, as this point of
  the parameter space is on the red line of
  Fig.~\ref{fig:Derrick_phase}. 
  (b) The energy as function of $R$, see Eq.~\eqref{erpol2}.
  In this figure $\kappa=0.974622$ and $1/e^2=0.001$ corresponding
  to $\tilde\kappa=1.73108$ and $1/\tilde{e}^2=0.000375625$. }
  \label{fig:hybridc}
\end{figure}
\begin{figure}[!htp]
  \centering
  \mbox{\subfloat[]{\includegraphics[width=0.49\linewidth]{{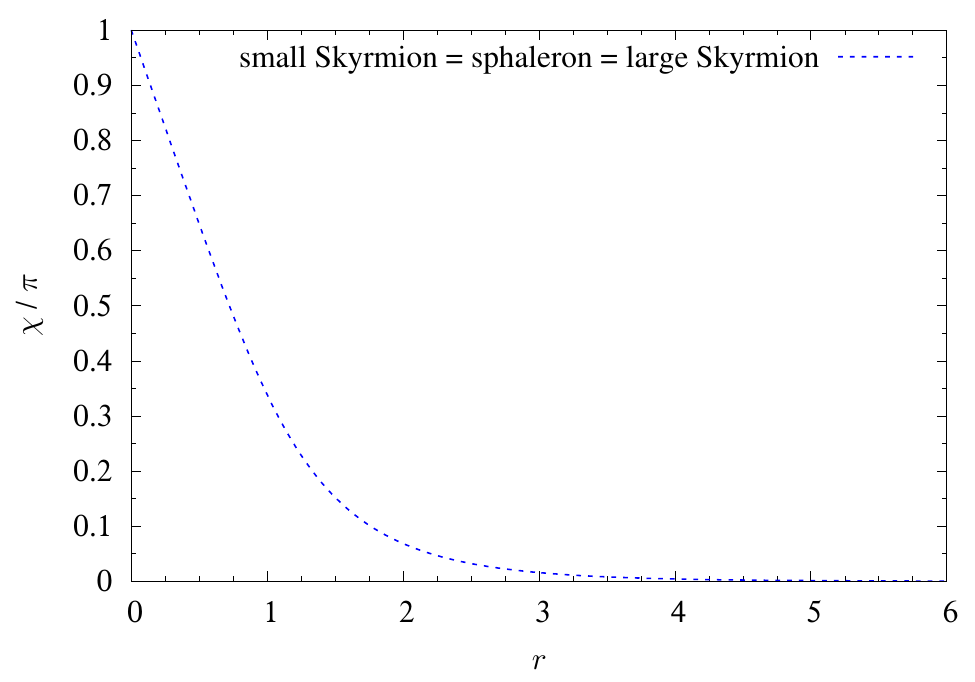}}}
    \subfloat[]{\includegraphics[width=0.49\linewidth]{{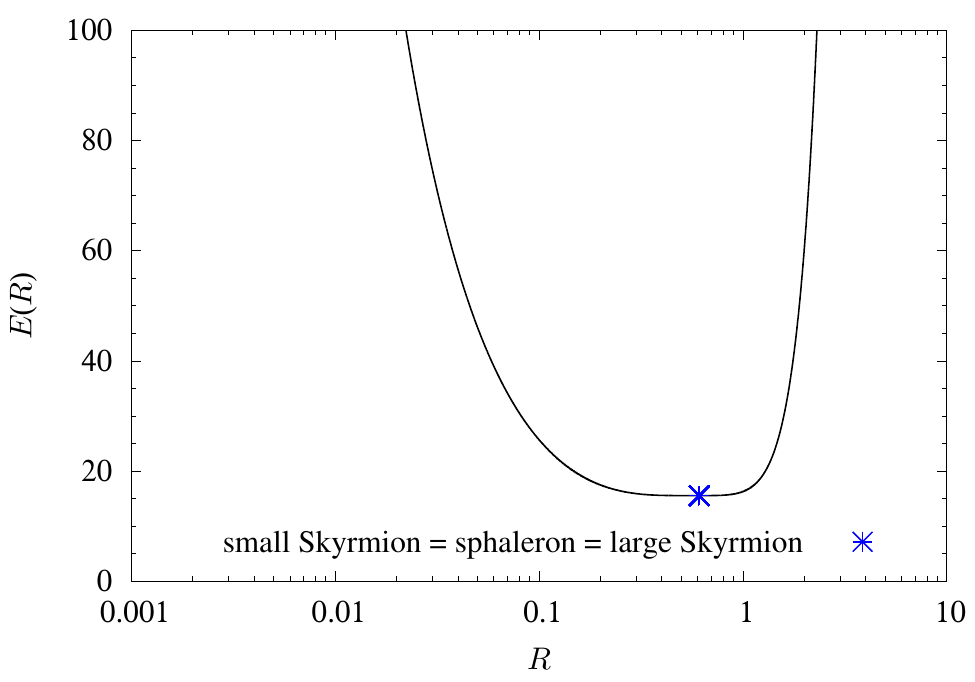}}}}
  \caption{(a) The solution in the magnetic-QCD hybrid Skyrme
  model at the triple point in parameter space corresponding to D in the
  phase diagram \ref{fig:Derrick_phase}.
  The small Skyrmion, the sphaleron and the large Skyrmion all have
  merged at this triple point. 
  (b) The energy as function of $R$, see Eq.~\eqref{erpol2}.
  In this figure $\kappa=0.948102$ and $1/e^2=0.0654157$ corresponding
  to $\tilde\kappa=\sqrt{8/3}$ and $1/\tilde{e}^2=1/36$. }
  \label{fig:hybridd}
\end{figure}
\begin{figure}[!htp]
  \centering
  \mbox{\subfloat[]{\includegraphics[width=0.49\linewidth]{{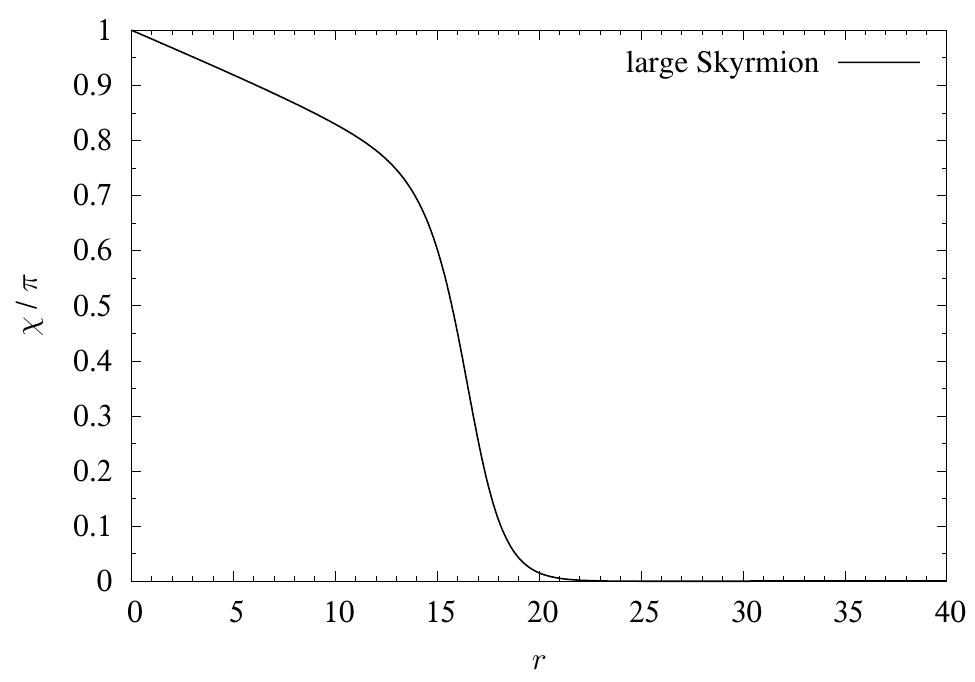}}}
    \subfloat[]{\includegraphics[width=0.49\linewidth]{{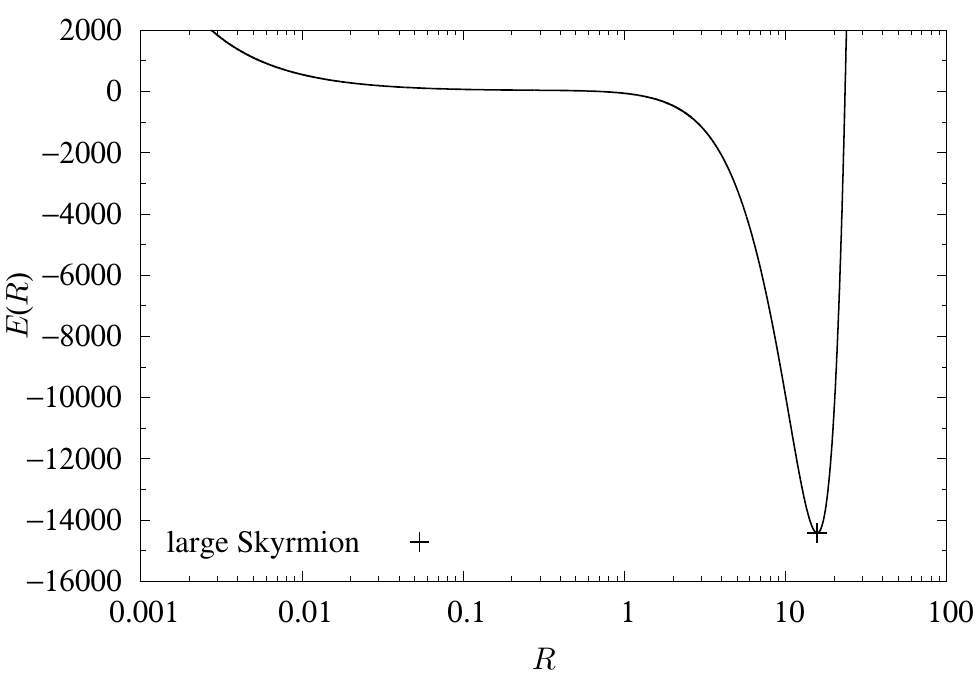}}}}
  \caption{(a) The solution in the magnetic-QCD hybrid Skyrme
  model at the point in parameter space corresponding to E in the
  phase diagram \ref{fig:Derrick_phase}: the large Skyrmion, which is stable.
  (b) The energy as function of $R$, see Eq.~\eqref{erpol2}.
  In this figure $\kappa=5$ and $1/e^2=1/25$ corresponding
  to $\tilde\kappa=6.37029$ and $1/\tilde{e}^2=0.00343907$. }
  \label{fig:hybride}
\end{figure}

For the stable and metastable solutions, we use the gradient flow
method for finding the numerical solutions.
This method is, however, not very effective at finding the unstable
solutions, for which we use the shooting method.
Knowing the size from the fixed points of the energy \eqref{erpol2}
makes it relatively easy to guess the shooting parameter $c$, where
$\chi=\pi-c r+\mathcal{O}(r^2)$.

As we can see from the phase diagram in Fig.~\ref{fig:Derrick_phase},
there are generally 5 different situations: in the green area there
are 3 different solutions, on the black and red line there are two
different solutions, whereas at the triple point and in the white
region, there is only a single solution.
For simplicity, we show only one example of each of the 5 different
cases.

Starting with the point A of the phase diagram
\ref{fig:Derrick_phase}, we illustrate the 3 different solutions in
Fig.~\ref{fig:hybrida} which correspond to the (metastable) small
Skyrmion, the (unstable) sphaleron and the (stable) large Skyrmion.
At point B of the phase diagram \ref{fig:Derrick_phase}, we show two
different solutions in Fig.~\ref{fig:hybridb} which correspond to the
sphaleron and the large Skyrmion; the small Skyrmion and the sphaleron
have merged into a single solution, as is shown with a black line in
the phase diagram \ref{fig:Derrick_phase}. 
At point C of Fig.~\ref{fig:Derrick_phase}, we show two different
solutions in Fig.~\ref{fig:hybridc} corresponding to the small
Skyrmion and the sphaleron; in this case the sphaleron and the large
Skyrmion have merged, leaving the small Skyrmion as the stable solution.
At point D of the phase diagram \ref{fig:Derrick_phase}, we show the
only existing
solution in Fig.~\ref{fig:hybridd}; this solution is at the triple
point of the phase diagram, indicating that all three solutions have
merged into a single stable Skyrmion.
At point E of Fig.~\ref{fig:Derrick_phase}, we show the only existing
solution in Fig.~\ref{fig:hybride}, which is a large Skyrmion and is
stable.

\subsection{Connecting to the Hopfion}

Analogously to the case without the Skyrme term
(Sec.~\ref{sec:magSkHopf}), the hybrid magnetic-QCD Skyrme model can
also be connected to Hopfions, which due to the Skyrme term will be
DMI-deformed Faddeev-Skyrme Hopfions.
When $\kappa:=0$ it is exactly the Faddeev-Skyrme model \cite{Faddeev:1975}.
The Derrick scaling of this theory (DMI-deformed Faddeev-Skyrme model)
is exactly the same as our hybrid magnetic-QCD Skyrme model.
We hence expect that the Hopfion model also contains both a \emph{sphaleron}
as well as a small Hopfion, in some parts of the parameter space,
similarly to what happened for the Skyrmions in
Sec.~\ref{sec:hybrid_Derrick_scaling}.

\section{Physical realization of higher-dimensional magnetic solitons}\label{sec:synthetic}

In the realm of quantum simulations for condensed matter phenomena,
the idea of ``synthetic dimensions'' \cite{Boada2012,Hazzard2023} has
recently become a valuable tool for replicating effects like
topological phases of matter \cite{Ozawa2019,Arguello-Luengo2024},
which arise in higher-dimensional physical systems.
This concept revolves around connecting certain degrees of freedom,
such as various internal states of atoms
\cite{Price2015,Ozawa2015,Price2022} or photons \cite{Yuan2018}, to
emulate the movement of a particle along an additional spatial
dimension. 
Lattice Hamiltonians allow for the simulation of higher-dimensional
topological models in lower-dimensional systems, though they require
reinterpretation of internal degrees of freedom to emulate an extra spatial
dimension.
Traditional methods, like embedding higher-dimensional lattices or
using topological pumps, are challenging or limited in capturing full
dynamics. 

\begin{figure}
    \centering
    \includegraphics[width=\linewidth]{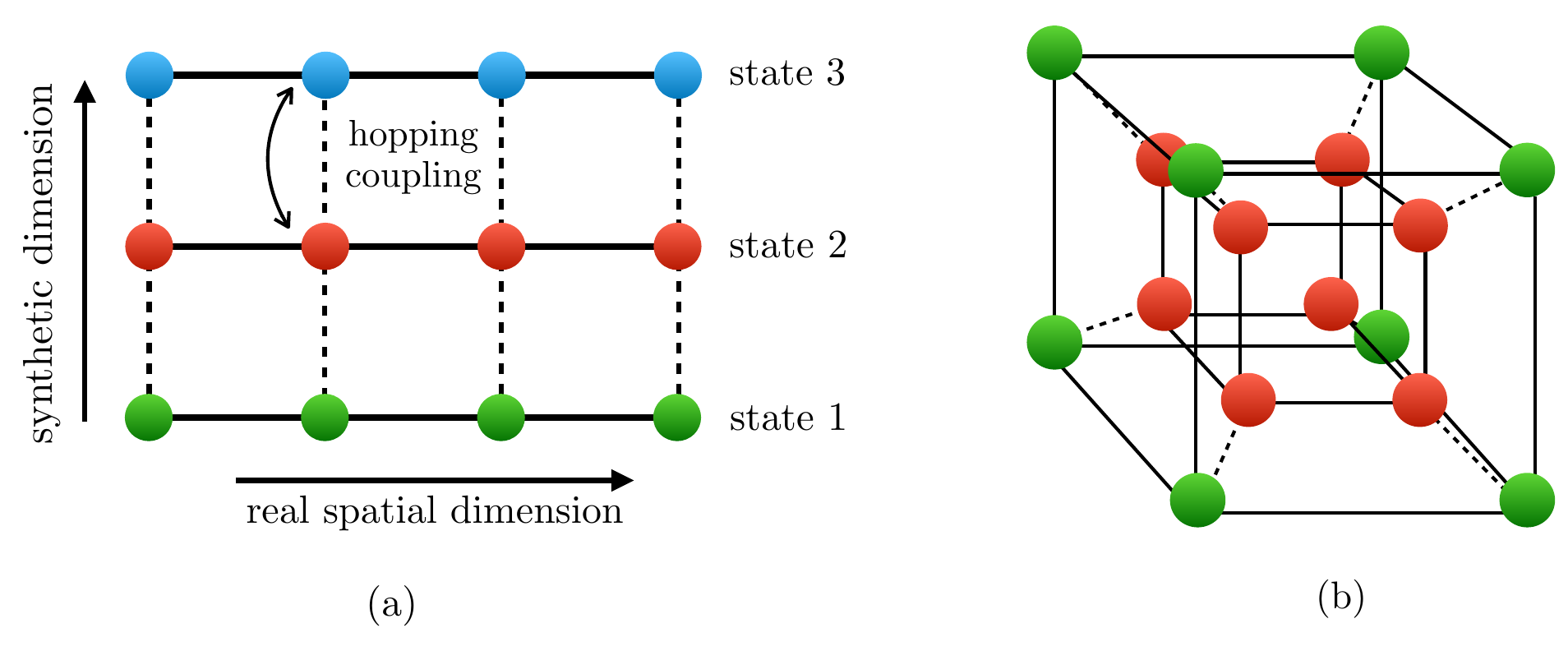}
    \caption{Synthetic dimensions transform internal states or
      intrinsic properties into an effective spatial dimension. (a) A
      2D discrete lattice model includes one real spatial dimension
      and one synthetic dimension composed of spin states. Hopping
      along the real dimension (solid lines) represents actual
      particle motion, while hopping along the synthetic dimension
      (dashed lines) corresponds to externally induced transitions
      between spin states. (b) A 4D hypercubic lattice illustrates a
      combination of real and synthetic spatial dimensions, showcasing
      how higher-dimensional structures can emerge in
      lower-dimensional systems.
    } 
    \label{fig:synthetic}
\end{figure}
In contrast, synthetic dimensions offer an innovative path for
lower-dimensional systems to explore higher-dimensional physics.
The key idea is to take a set of unconnected internal states within a
system and re-imagine them as lattice sites along a spatial
dimension.
By introducing specific external couplings, particles can effectively
``hop'' along this synthetic dimension, mimicking movement across a
real lattice \cite{Celi2014} as shown pictorially in Fig.~\ref{fig:synthetic}.

What remains to be specified is the nature of the internal states of
the system which will effectively act as an extra spatial
dimension. For magnetic Hopfions or Skyrmions, the spin-orbit coupling
(SOC) is especially promising because it influences chirality and
texture stability, giving rise to complex configurations that could
correspond to synthetic 4D behavior.
We leave the search for the exact physical system that could be a
candidate for providing the synthetic dimension for our
higher-dimensional version of the DM interaction for future work. 

\begin{figure}[!htp]
  \centering
  \includegraphics[width=0.6\linewidth]{{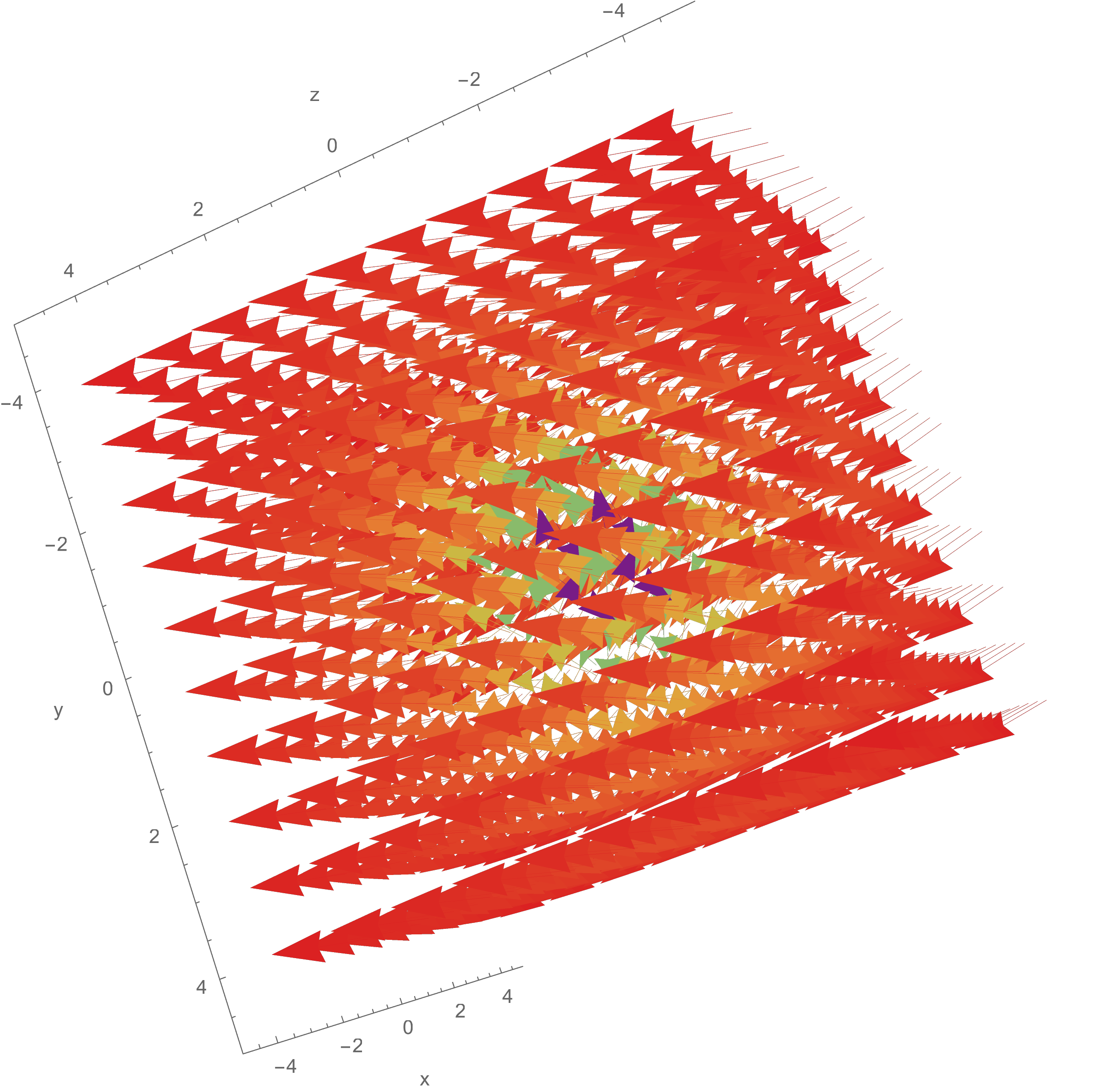}}
  \caption{Illustration of the magnetization vector in an example
    where the physical dimensions are mapped as
    $\bm{n}=(n^x,n^y,n^{\rm synth},n^z)$, with $n^{\rm synth}$ representing
    the component of the would-be magnetization vector in the synthetic dimension.
  }
  \label{fig:synth_proj}
\end{figure}
In Fig.~\ref{fig:synth_proj} we illustrate an example of a Skyrmion
when the $n^3$ element of the 4-dimensional magnetization vector
correspond to the synthetic dimension.

\section{Conclusion}\label{sec:conclusion}

In this paper, we have generalized the Dzyaloshinskii-Moriya
interaction for a 3D magnetization vector to a 4D magnetization vector
-- hence a higher-dimensional version of the DMI.
Using the property that it is first order in derivatives and must be
nonvanishing, we deduced that it must be contracted with a tensor that
is antisymmetric in $\Og(4)$-indices and bears a 3D spatial index.
Considering furthermore the simplified case of $\SO(3)_{\rm diag}$ invariance, we
are able to reduce the DM term to just two invariant structures, that
we denote the $\alpha$ and $\beta$ parts.
It turns out that the $\beta$-part is irrelevant for the spherically
symmetric case and only the $\alpha$-part stabilizes the solitons.
Due to the higher dimension of space, there are in fact two solitons
with just the DMI and a potential: the Skyrmion and a sphaleron,
where the latter is an unstable and smaller solution of the model.
Upon reduction to the Hopfion model, the $\beta$-part nevertheless
plays an important role.
Finally, we consider including the Skyrme term, thus considering a
magnetic-QCD hybrid Skyrme model that turns out to host a large and a
small Skyrmion as well as a sphaleron -- at least in a small region of
parameter space of the model.
The large Skyrmion is stable, the small Skyrmion is metastable and the
sphaleron is unstable.
The phase diagram is quite rich and there exist lines where the
different types of solutions coalesce and in the remaining part of the
parameter space of the model, there exists only one solution --  the
Skyrmion.

We have contemplated that by using a synthetic extra dimension, it
would be possible to construct our higher-dimensional magnetic
Skyrmion or perhaps the hybrid Skyrmion in a physically realizable
condensed matter system.
The details of whether the exact Hamiltonian corresponding to our
model can be constructed, still needs to be worked out.

We have focused on spherically symmetric solitons in this paper.
Using the spherically symmetric hedgehog Ansatz, we found that
$\beta$-part of the $\SO(3)_{\rm diag}$-invariant DM term
\eqref{eq:Theta_inv_standard} vanishes and so we effectively only used
the $\alpha$-part of the $\SO(3)_{\rm diag}$-invariant DM term.
Deformed solitons, however, may feel the presence of the $\beta$-part
of the DM term, which in principle could lower the mass of the
soliton.
We leave the exploration of such a possibility for future work.

We also considered a connection to the Hopfion in chiral magnets, and
although there exists a higher-dimensional DM term and a potential with
a limit that connects to the Hopfion model, the interpolation is not
expected to be smooth. That is, the topology changes, sending the
masses of the nontrivial $\pi_3(S^3)$ solitons to infinity and the
nontrivial $\pi_3(S^2)$ solitons appear first as metastable states and then finally
become stable at the end point of the limit.
A smooth interpolation is thus desirable and will be left for future work.

In summary, this work presents an initial study of a generalized model
for higher-dimensional magnetic Skyrmions, revealing rich solitonic
physics.

\subsection*{Acknowledgments}
S.~B.~G.~thanks the Outstanding Talent Program of Henan University for
partial support.
The work of S.~B.~G.~is supported by the National Natural Science
Foundation of China (Grant No.~12071111) and by the Ministry of
Science and Technology of China (Grant No.~G2022026021L).
The work of S.~B. is supported by the INFN special research project
grant ``GAST'' (Gauge and String Theories).

\bibliographystyle{JHEP}
\bibliography{biblio}
\end{document}